# Complex polar superstructure controlled thermal conductivity in ferroelectric $PbTiO_3/SrTiO_3$ superlattices.

Noa Varela-Domínguez,[1] Marcel S. Claro,[1] Araceli Gutiérrez-Llorente,[2] Eric Langenberg,[3] Xinxin Hu,[4] Núria Bagués,[5] Anthony Edgeton,[6] Chang Beom-Eom,[6*] Jordi Arbiol,[4,7] José Santiso,[4*] Francisco Rivadulla.[1*]

[1]CiQUS & Departamento de Química-Física, Universidade de Santiago de Compostela, 15782 Santiago de Compostela, Spain.

[2]Escuela Superior de Ciencias Experimentales y Tecnología, Universidad Rey Juan Carlos, Madrid 28933, Spain.

[3]Departament de Física de la Matèria Condensada & Institut de Nanociència i Nanotecnologia (IN[2]UB), Universitat de Barcelona, 08028 Barcelona, Spain.

[4]Catalan Institute of Nanoscience and Nanotechnology (ICN2), CSIC and BIST, Campus UAB, Bellaterra, Barcelona, Catalonia, 08193 Spain.

[5]ALBA Synchrotron Light Facility, Carrer de la Llum 2-26, Cerdanyola del Vallès, Barcelona 08290, Spain.

[6]Department of Materials Science and Engineering, University of Wisconsin-Madison, Madison, WI 53706, USA

[7]ICREA, Pg. Lluís Companys 23, 08010 Barcelona, Catalonia, Spain

*e-mail: f.rivadulla@usc.es
*e-mail: jose.santiso@icn2.cat
*e-mail: ceom@wisc.edu

**Abstract.** Integrating epitaxial thin films of ferroelectric $PbTiO_3$ and paraelectric $SrTiO_3$ into artificially layered periodic superlattices provides a unique platform for tuning strain, depolarization, and interfacial/surface energies, thereby accessing a rich phase diagram of topological polar structures (skyrmions, vortices, merons, or sinusoidal waves) and superstructures (polar supercrystals). Here we show that the 3D arrangement of polar vortices in a supercrystal suppresses thermal conductivity ($\kappa$) of PTO/STO superlattices (SLs). The temperature dependence of $\kappa$ reflects the evolution of the polar superstructure, as determined by X-ray diffraction and transmission electron microscopy. The comparison with other SLs suggests that the 3D arrangement is crucial for controlling thermal conductivity beyond the usual interfacial scattering. Moreover, we observed an unexpected reduction in thermal conductivity with increasing superlattice thickness, a phenomenon reminiscent of phonon-wave Anderson localization. Our results show that complex polar superstructures can be useful active elements for modulating heat transport in technologies where control over heat dissipation is critical.

**Introduction.** The interplay among strain, gradient and depolarization energies governs the type and distribution of domains in ferroelectric materials. Epitaxial ferroelectric/dielectric superlattices offer a unique platform to independently tune these energy contributions through layer thickness, composition, and epitaxial interface engineering. This tunability can drive the system into regions of the phase diagram where multiple structural or polar phases become nearly degenerate, leading to the emergence of complex, non-trivial polarization patterns that minimize the total free energy.[1]

For instance, the periodicity of $(PTO)_n/(STO)_n$ superlattices (SLs) controls the electrostatic coupling among the ferroelectric PTO layers through the thickness of the intermediate, and nominally paraelectric, STO. Growing these SLs on top of $AScO_3$ (A= Dy, Gd, Sm or Nd) substrates imposes a slight tensile strain, which results in periodic arrays of clockwise and counterclockwise vortex-antivortex pairs in partially decoupled PTO layers (n=10-20 unit cells); smooth sinusoidal patterns along the in-plane direction for partially coupled (n=7-11 u.c.); and the stabilization of polar antivortices within the STO layers at n=4, demonstrating that these topological polarization textures can be extended to the whole SL in the coupled regime.[2–6] In this regard, mesoscopic, long-range ordered structures exhibiting three-dimensional periodicity of these polarization textures were observed by Stoica *et al*.[7] In their case, this polar supercrystal results from a spontaneous organization of polar vortices in a $(PbTiO_3)_{16}/(SrTiO_3)_{16}$ superlattice grown on $(110)_o$ $DyScO_3$ (DSO), after ultrafast laser irradiation. The generation of free carriers that accumulate at the PTO/STO interface partially screens and reduces the depolarization field, enabling the stabilization of this long-range polar configuration. The vortex topology of the supercrystal can be manipulated by temperature and electric fields, showing a rich phase diagram.[8]

On the other hand, similar SL periodicities grown on $SrTiO_3$, which imposes a mild compressive strain, result in the stabilization of arrays of polar skyrmions along the PTO layers.[9] These nanoscale polar bubbles are similar to magnetic skyrmions, with a continuous rotation of the electric polarization in a hedgehog texture with a diameter $\approx$8 nm, and a periodicity along the in-plane direction that can be modulated by the PTO thickness.[10]

Thus, tuning the periodicity, epitaxial strain, and growth conditions of PTO/STO SLs allows the stabilization of a wide variety of competing polar configurations; their nanoscale dimensions and periodicity, along with the possibility of reconfiguring/erasing them with an electric field or temperature opened the possibility of designing new types of devices for non-volatile topological memories, race-track shift registers, reservoir computing, etc.[11,12]

Although polar textures are intrinsically coupled to phonons, their influence on the heat propagation in ferroelectric superlattices has remained largely unexplored. This is particularly noteworthy given that ferroelectric domain walls have been identified as strong phonon scatterers,[13] and that complex polar topologies may impart chirality to phonons, thereby enabling control over heat flow via phonon Hall effects.[14]

In superlattices, mid- to long-wavelength phonons are able to propagate coherently across the periodic interfaces, providing the dominant contribution to thermal transport.[15–17] However, their role is dictated by the superlattice periodicity, which is predetermined during growth and cannot be modified once the structure is fabricated, limiting post-synthesis control over heat conduction. In this context, reconfigurable polar structures with adjustable periodicities introduce a new degree of freedom, suggesting a previously unexplored route toward the active modulation of thermal conductivity.

To fill this gap, here we report the thermal conductivity across $(PTO)_n/(STO)_n$ SLs with different periodicities (n, in unit cells) and strain. Notably, the presence of vortex arrangements through the SL leads to an anomalous reduction in thermal conductivity as SL thickness increases, resembling phonon-wave Anderson localization.[18,19] More importantly, the long-range ordering of vortices into a supercrystal produce a dramatic suppression of thermal conductivity, which can be partially recovered by E-field or temperature destabilization of the polar structures. These results highlight that polar topologies provide an additional and reversible degree of control over phonon transport. We conclude that complex polar textures can serve as functional elements for thermal management in oxide superlattices, with potential implications for thermoelectric performance and heat dissipation in nanoscale devices.

**Results and discussion.**

For this work we fabricated a series of $[(PTO)_n/(STO)_n]_y$ SLs on (001) STO and $(001)_{pc}$ DSO substrates, systematically varying the periodicity (n=4, 15, 25) and total number of repetitions (y=6 to 60) to achieve comparable total SL thicknesses.

Growing on $(001)_{pc}$ DSO imposes a tensile (compressive) strain on STO (PTO), which results in a very rich polar phase diagram as a function of the periodicity of $[(PTO)_n/(STO)_n]$ SLs:[5] arrangements of $a_1/a_2$ domains in the PTO layers for periods up to n≈6-8 unit cells; flux closure domains in the large-period regime (n≳25 unit cells), when the relaxation of elastic energy and the expansion of domain size makes the depolarization energy to dominate; and a region of complex polar vortices which compete with a continuous sinusoidal oscillation of the polarization along the PTO layers, when the thickness of the PTO and STO layers is ≈10-20 u.c.[20] In this intermediate periodic thickness, charge accumulation at the interfaces by optical excitation results in the stabilization of a supercrystal, with 3D polar and strain periodicities.[7,8]

The emergence of these complex polarization structures in the $[(PTO)_{15}/(STO)_{15}]_{15}$ SLs grown on DSO is confirmed through the high-resolution STEM-HAADF analysis of cross-section lamellae shown in Figure 1.

The STEM image obtained with atomic column resolution along the [010] zone axis of the SL (Figure 1 a), and the polar displacement map obtained from this image (Figure 1 b-c) shows a collective sinusoidal arrangement of the polarization along the PTO layers, with a lateral periodicity of ≈9.8 nm, while the STO layer remains paraelectric. This periodicity is consistent with previously reported polar vortex lattices in PTO/STO superlattices of the same periodicity.[21] For this reason, we will adopt this nomenclature hereafter.

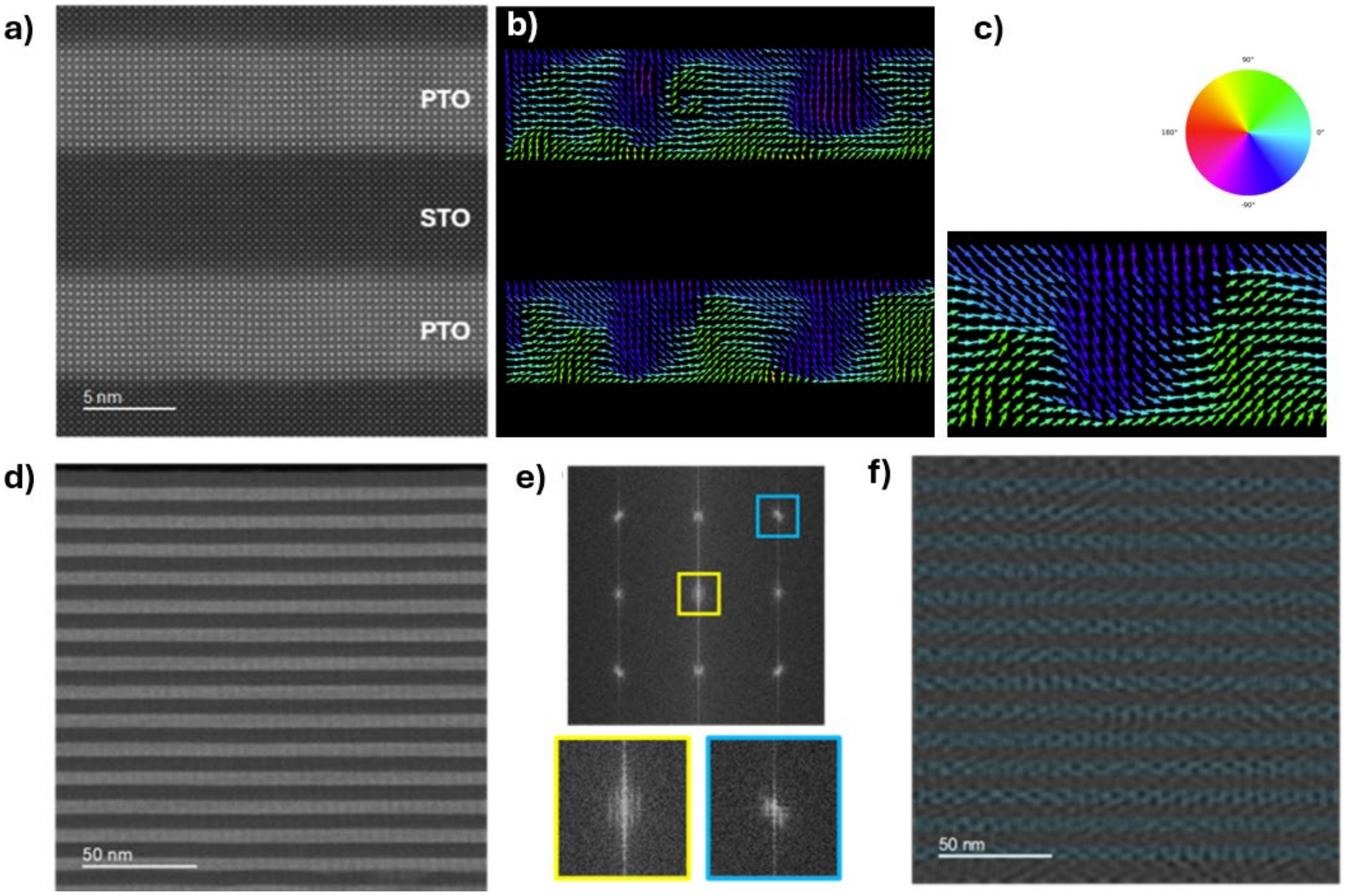

**Figure 1: Microstructural characterization of the SL and polar displacement maps.** a) High-resolution STEM-HAADF image displaying two consecutive PTO slabs of the $[(PTO)_{15}/(STO)_{15}]_{15}$ SL on DSO; $\Lambda_{SL}$= 12 nm in the vertical direction. b) Polar displacement map calculated from previous image and c) magnification of the polar map highlighting the wavy polar arrangement along the PTO slab; the color wheel defines the mapping between the vector direction and the color hue, corresponding to an specific angle. d) Low magnification image of the full SL, along with the Fast Fourier Transform (FFT) pattern of the whole image. e). The insets of (0,0) and (1,1) spots show multiple satellites, corresponding to a complex superstructure in/out of plane. f) Inverse FFT after including all superstructure satellites of (h,k) reflections up to second order, except zero order main reflections, and filtering out non-significant background regions. The contrast pattern in the image reveals the periodic oscillation of polarization along the PTO layers, as well as their correlation between consecutive PTO layers in some parts of the image.

A low magnification HAADF image of the SL oriented in the [010] zone axis is displayed in Figure 1 d), along with its FFT (Figure 1 e). Apart from the intense spots in the average structure of the SL, there is a subtle superstructure evidenced by multiple satellites around the main reflections. The selected regions in the image show the detail around two characteristic spots (0,0) and (1,1) to demonstrate the complex satellite pattern. The image of Figure 1 f) was obtained by filtering the selected areas of the satellite peaks in the FFT and applying inverse Fourier transform. The filtered image shows a clear contrast that evidences the formation of an alternating periodic pattern along the PTO slabs, with a periodicity between ≈10.5-11.2 nm.

Importantly, some regions of the image present a coupled pattern between adjacent PTO layers, across the paraelectric STO spacer. In some areas the coupling is antiphase, while in others it is in-phase. However, we were unable to identify extended regions with a perfectly defined coupling between adjacent PTO slabs. Also, additional atomic resolution HAADF STEM observations of the same lamella and after further FIB-thinning, removed the wavy pattern completely (see Supporting Information). This demonstrates/highlights the extreme sensitivity of these highly complex polarization patterns to the delicate balance between elastic strain and electrostatic charge at the interfaces, which are severely affected by sample preparation.[22]

Therefore, the safest (and non-invasive) way to probe the intrinsic nature and stability of such periodic polar superstructures is through the temperature dependence of high-resolution X-ray reciprocal space maps (RSM) of the unprocessed samples. The results for $[(PTO)_{15}/(STO)_{15}]_y$ are shown in Figure 2; similar X-ray experiments of the $[(PTO)_4/(STO)_4]_y$ and $[(PTO)_{25}/(STO)_{25}]_y$ SLs are discussed in the Supporting Information and confirm the presence of $a_1/a_2$ and flux closure domains, respectively.

Symmetric $\omega/2\theta$ scans (Figure 2 a) show multiple superlattice satellite peaks around the $002_{pc}$ reflection of DSO, confirming the long-range structural order. From the position of the (002) SL reflections and the zeroth-order signal, we can estimate the out-of-plane lattice parameter of the SL c= 3.933(2) Å. Similar analysis for the other SLs yields c= 3.919(3) Å for n=4 and c= 3.94(5) Å for n= 25, indicating a systematic expansion of the c-axis with increasing period thickness (see Supporting Information). The superlattice periodicity determined from the angular separation between the satellite peaks is $\Lambda_{SL}\approx$11.63 nm, in good agreement with the designed nominal periodicity ≈11.7 nm for $[(PTO)_{15}/(STO)_{15}]$ and the observations by HAADF STEM (Figure 1). This serves as an internal standard for determining other periodicities, as discussed below.

In addition to the SL satellites, a series of intermediate low-intensity reflections, labelled as $V_{\pm n}$ in Figure 2 a) are observed. These exhibit a characteristic out-of-plane parameter c= 3.972(4) Å and vertical periodicity $\Lambda_z\approx$ 11.86 nm. These signals are absent in the other SLs with shorter or longer periods (see supporting information) and match the periodicity of the sinusoidal modulation of the polarization observed in Figure 1 b,f).

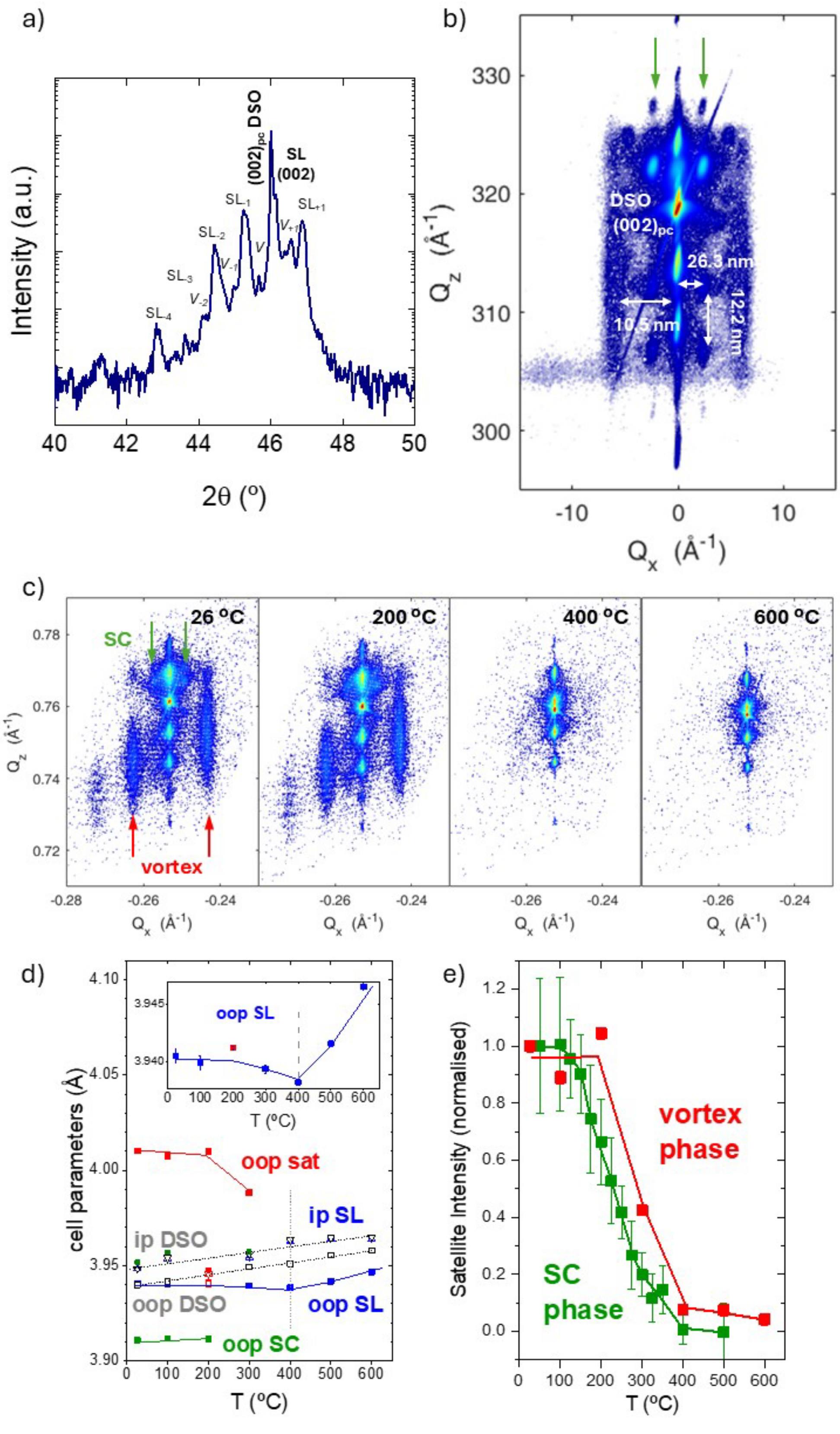

a)
Intensity (a.u.)
2θ (º)
DSO
(002)pc
SL
(002)
SL-4
SL-3
SL-2
SL-1
SL+1
V-2
V-1
V
V+1
b)
Qz (Å-1)
Qx (Å-1)
DSO
(002)pc
26.3 nm
10.5 nm
12.2 nm
c)
26 ºC
200 ºC
400 ºC
600 ºC
SC
vortex
Qz (Å-1)
Qx (Å-1)
d)
cell parameters (Å)
T (ºC)
oop SL
oop sat
ip SL
ip DSO
oop DSO
oop SC
e)
Satellite Intensity (normalised)
T (ºC)
vortex
phase
SC
phase

**Figure 2: X-ray analysis of the SLs.** a) $\omega/2\theta$ scans and b) RSM of the $[(PTO)_{15}/(STO)_{15}]_{15}$ SL around the $(002)_{pc}$ reflection of DSO. The vertical arrows in b) identify the reflections from the supercrystal structure (see text). c) RSM of the $(103)_{pc}$ reflection at different temperatures showing the evolution of the characteristic signals of different polar textures (supercrystal and vortex domains). d) Temperature evolution of the in-plane (ip) and out-of-plane (oop) cell parameters of the SL, DSO substrate, SC and vortex phase (sat). Inset shows an enlarged view of the oop cell parameter measured for the main SL peaks evidencing the change in slope characteristic of the PTO ferro-paraelectric transition at $T_c$= 400 ºC. e) Intensity of the SC and vortex phase satellites up to 600 ºC (note that SC phase starts reducing at lower temperature than vortex phase).

The periodic arrangement of the polar structure is further confirmed by the presence of satellite peaks in the horizontal direction of the X-ray reciprocal space map (RSM; indicated with red arrows in Figure 2 c), with a spacing along $Q_x$ corresponding to a lateral periodicity of $\Lambda_x$= 10.5 ±0.2 nm at room temperature, corresponding to the vortex phase arrangement.

On the other hand, the RSMs show additional features which consist of a separate collection of well-defined and narrow satellite peaks, with a periodicity along the vertical direction matching that of the SL, while their horizontal periodicity, measured along the $[100]_{pc}$ in-plane direction of the DSO substrate is notably larger than that the vortex phase, with $\Lambda_x$= 26.3 ±0.3 nm. Notably, the $a_1/a_2$ microstructure in the PTO films presents satellites along the [110]/[1-10] directions in reciprocal space, while the satellites observed in Figure 2 b) align along [100] direction, suggesting a different structural arrangement.

These two distinct length scales (≈10 nm and ≈26 nm) indicate the presence of multiple lateral periodicities in the sample, consistent with a SL that displays both primary vortex arrays and secondary superstructure.

These reflections are characteristic of a stable long-range ordering of nanoscale polar vortices into a supercrystal (SC) phase, which emerges from the mixture of the $a_1/a_2$ and vortex domains at intermediate thickness periodicity.[7] However, unlike previous works, we stabilize this phase without the need of optical accumulation of charges at the PTO/STO interfaces. The specific growth conditions used for our SLs seem to tune the competing energies in a way that favors stabilization of the SC phase.

To address the thermal stability of the supercrystal we measured the RSM up to 600 ºC, monitoring the evolution of the most intense satellites as a function of temperature (Figure 2 c).

As shown in Figure 2 d), the in-plane (ip) lattice parameters of the SC phase match that of substrate, consistent with a fully coherent structure, while the out-of-plane (oop) values remain almost constant at about 3.911 Å from room temperature up to 200 ºC. Above this temperature, the superstructure satellites vanish.

The evolution of the intensity of both the vortex and SC phases with temperature is compared in Figure 2 e). For the vortex satellites, the intensity remains stable up to 200 ºC, then decreases, vanishing at about 400 ºC. In contrast, the intensity of the main satellites for the SC phase is maintained up to 100 ºC, then gradually decreases disappearing at 300 ºC. This subtle difference indicates different prevalences depending on the temperature range: i) from room temperature to ≈150 ºC the SL stabilizes a mixed state of polar vortex and a long-range correlated supercrystal; ii) from ≈150 ºC to ≈300 ºC the SC disappears, prevailing the signal of the vortex phase; iii) above ≈300 ºC only the vortex phase is still present; iv) beyond ≈400 ºC the SL undergoes a transition into the paraelectric phase.

The temperature dependence of the thermal conductivity of the SL mirrors the thermal stability of each of these phases; Figure 3 a). In particular, the thermal conductivity remains very low and nearly constant at constant at ≈2.25 W $m^{-1}K^{-1}$ below 150 ºC, corresponding to the SC phase, and increases only slightly above this temperature, during the vortex phase. Finally, once the polar structures are erased and the SL becomes paraelectric, the thermal conductivity increases rapidly, reaching ≈3 W $m^{-1}K^{-1}$ at 550 ºC. This value is similar to the maximum thermal conductivity we measured in $[(PTO)_4/(STO)_4]_y$ and $[(PTO)_{25}/(STO)_{25}]_y$ SLs, where these polar structures are absent. These results demonstrate that the complex 3D arrangement of polarization is the key element controlling the thermal conductivity in these SLs.

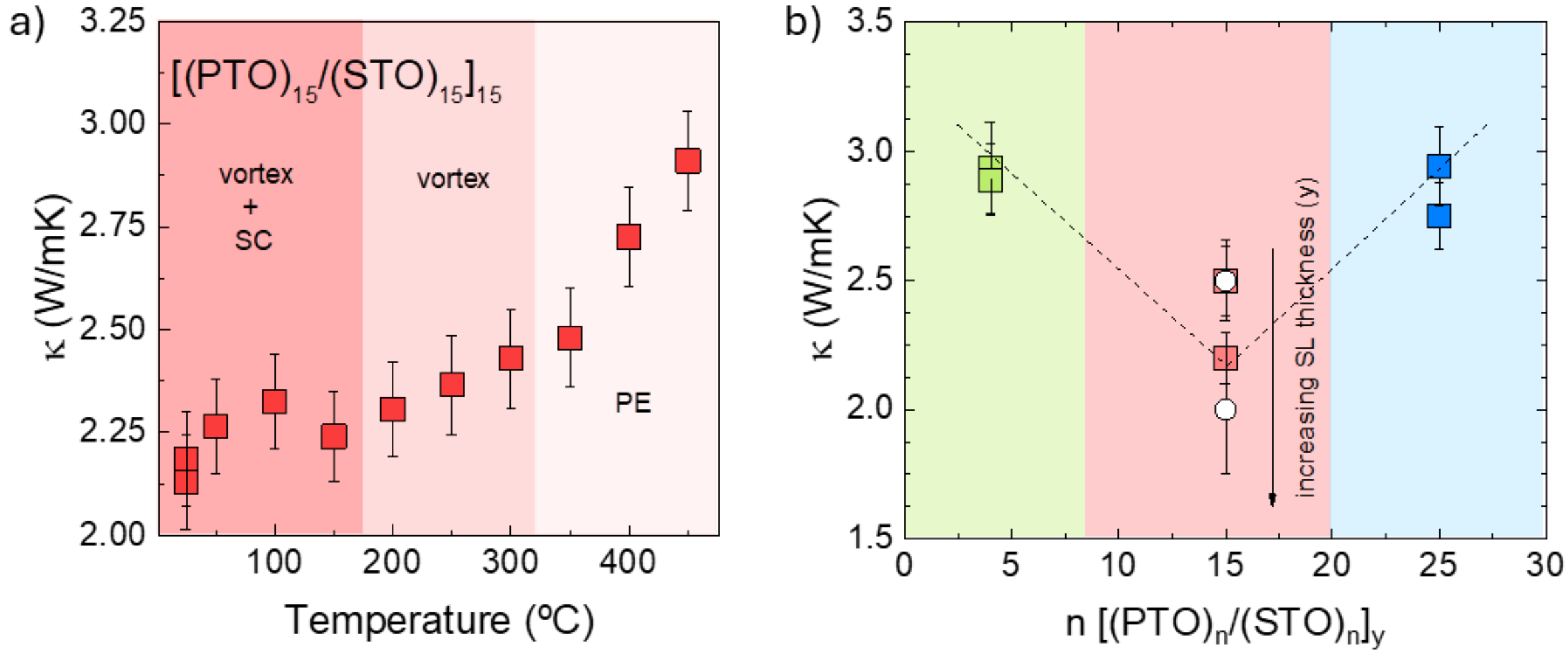

**Figure 3: Thermal characterization of the SLs**. a) Temperature dependence of the SL $[(PTO)_{15}/(STO)_{15}]_{15}$ on DSO, showing the different polarization regions identified through STEM and X-ray diffraction analysis. b) Thermal conductivity at room temperature for $[(PTO)_n/(STO)_n]_y$ on DSO SLs with different periods and total thickness (y). There is a reduction of the thermal conductivity at the n=15 period thickness, where the complex polarization patterns are identified. Note also the decrease of thermal conductivity with the total thickness of the SL. The results correspond to samples of 120 nm and 170 nm of the same periodicity, deposited on DSO (filled squares) and STO (open symbols; see also Figure 5 below).

Noteworthy, the effects reported in Figure 2 a-c) and Figure 3 a) are reversible, as evidenced by the reappearance of the SL peaks in the RSM and the reduction in thermal conductivity $\kappa$, upon cooling back to room temperature (Supporting Information).

On the other hand, we observed an unexpected reduction in the thermal conductivity of the SL with increasing total thickness (Figure 3 b). This effect is also observed in the $[(PTO)_{15}/(STO)_{15}]_y$ superlattices deposited on STO (open symbols in Figure 3 b) at the same periodicity. In this case, a complex periodic rotation of the polarization has been identified (Figure 4), although limited to the individual PTO layers, lacking 3D ordering. However, the reduction in thermal conductivity with increasing SL thickness is of the same order as that observed in the SLs grown on DSO, pointing to a general phenomenon occurring at intermediate periodicities.

Importantly, a similar reduction in thermal conductivity $\kappa$ with increasing total thickness has been reported in GaAs/AlAs SLs containing nanoparticles randomly distributed at the interfaces, as well as in aperiodic Si/Ge superlattices.[18,19] These are two distinct forms of disorder, which introduce destructive interferences and Anderson-like localization of phonon waves.

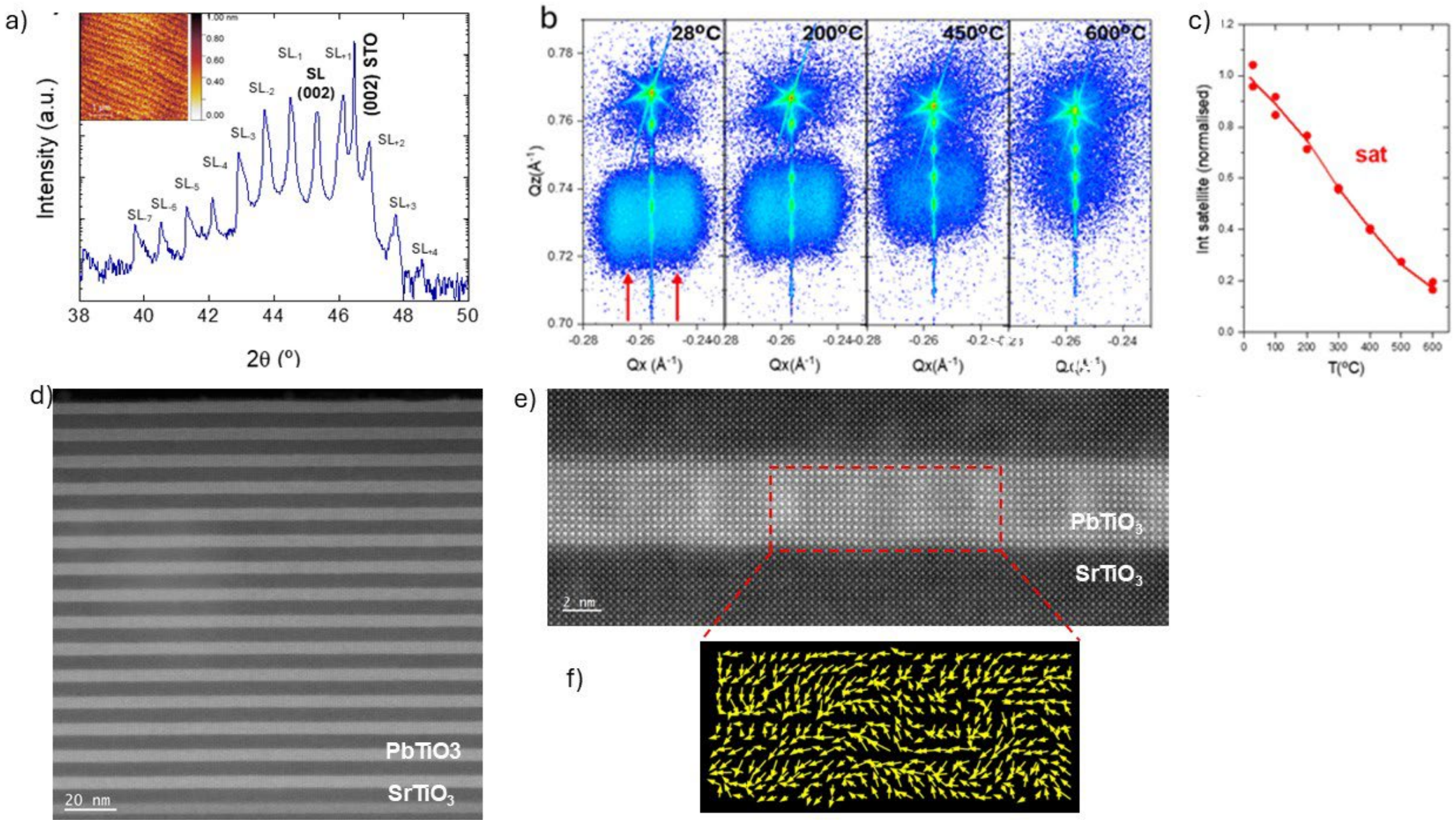

**Figure 4: Structural characterization of the $[(PTO)_{15}/(STO)_{15}]_{15}$ on STO SL.** a) $\omega/2\theta$ scan around 002 STO, with the SL peaks labeled; b) RSMs around the 103 STO measured at different temperatures. The red arrows indicate the lateral satellite arrangement of a periodic phase. c) Temperature dependence of the integrated intensity of the lateral satellite peaks of the vortex phase, normalized to RT value. d) Low magnification STEM-HAADF image of the full SL, and e) higher magnification image displaying the intensity modulations along the PTO layers, which suggest a periodic arrangement of the polarization. The polar displacement vectors calculated from the image (individual dipole vectors indicated with yellow arrows), show the complex, sinusoidal-like variation of the polarization along the PTO slab. (f)

In our case, however, the presence of X-ray SL peaks up to high order and the STEM images prove the near-perfect periodicity and minimal interfacial mixing. Moreover, the size effect is maximized at the periodicities where complex polar topologies, like vortex waves or supercrystals, are observed. This suggests that it is the partial random displacements of the polarization at the PTO/STO interfaces (see Figures 1 c and 4 f), what introduces a similar source of disorder, suppressing thermal conductivity as total thickness increases.

Finally, the temperature dependence of the thermal conductivity of the $[(PTO)_{15}/(STO)_{15}]_{15}$ SL on STO is shown in Figure 5. To enable the application of an electric field, a 5 nm $SrRuO_3$ buffer layer was deposited prior to the SL growth to serve as a bottom electrode. As shown in Figure 5 a), κ(T) presents a slight upturn above ≈430 ºC, coinciding with a significant reduction in the intensity of the lateral satellite peaks associated with the vortex phase, as previously detailed in Figure 4 c).

More importantly, the application of an electric field results in a large and fully reversible increase of the thermal conductivity, reaching values of 2.8-3 W $m^{-1}K^{-1}$ (Figure 5 b). It should be noted, however, that due to non-ideal insulation properties, the samples exhibit a finite leakage current. Consequently, Joule heating may occur within the SL, suggesting that the observed enhancement could arise from the combined effect of the applied electric field and thermal fluctuations.

Nevertheless, these findings demonstrate a significant advancement, given the reconfigurable nature of these polar structures and superstructures, which can be actively tuned via temperature or electric fields. Thus, this mechanism introduces a novel degree of freedom for thermal transport engineering in oxide-based superlattices.

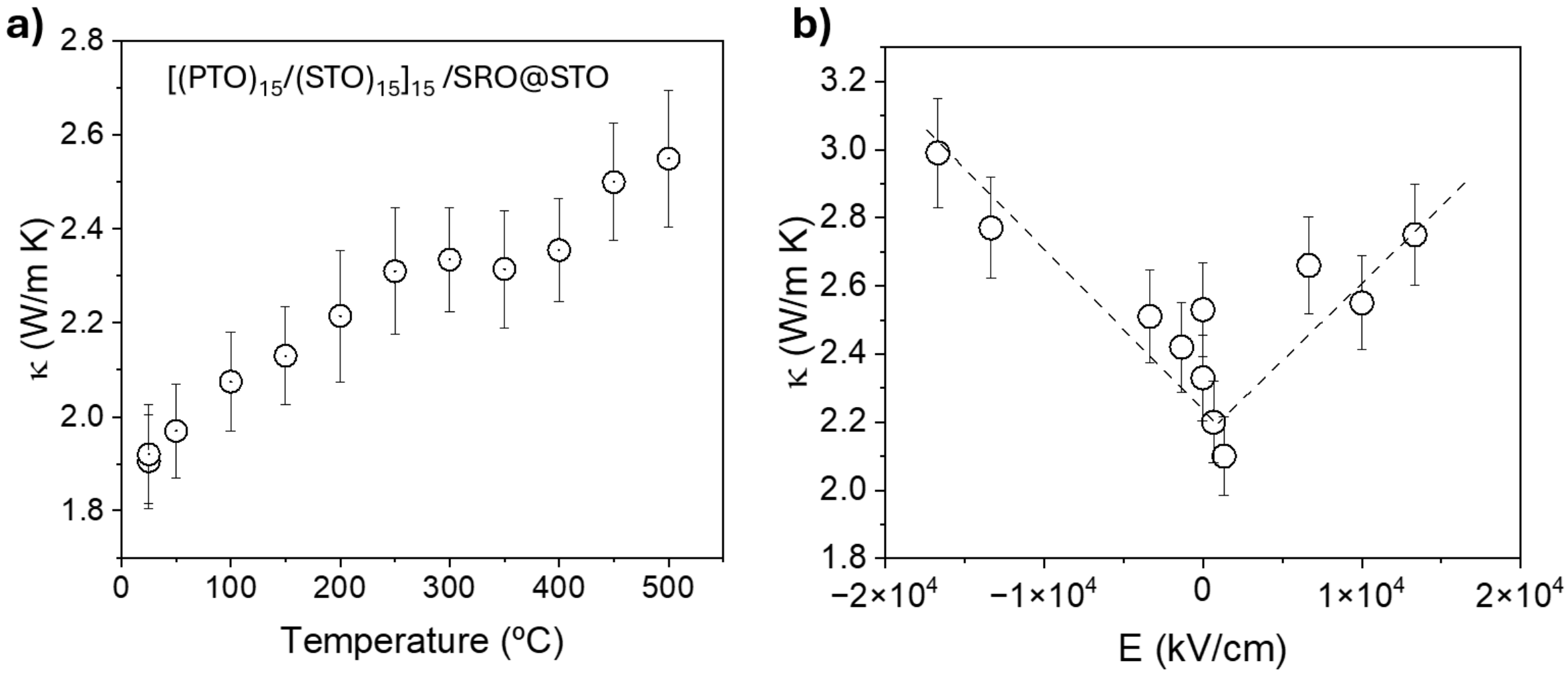

**Figure 5: Thermal conductivity of the $[(PTO)_{15}/(STO)_{15}]_{15}$ on STO SL.** a) Temperature dependence of the SL $[(PTO)_{15}/(STO)_{15}]_{15}$ on STO, showing. b) Effect if the electric field on the thermal conductivity of the SL at room temperature. In this case, the SL is grown on a buffer layer of $SrRuO_3$, which was used as the bottom electrode.

## Conclusions

We have shown that polar textures in (PTO) /(STO) SLs determine their thermal transport, revealing a new path for controlling phonon propagation in complex oxides. By systematically varying the superlattice periodicity and strain, we demonstrated that vortex lattices induce an unexpected reduction in thermal conductivity as the superlattice thickness increases, a phenomenon reminiscent of phonon-wave Anderson localization. Moreover, the emergence of long-range flux-closure ordering into a supercrystal state results in a dramatic suppression of thermal conductivity.

These findings demonstrate that polar topologies act not merely as passive structural features but as active elements for modulating heat transport. Our experiments show the possibility to reversibly tune thermal conductivity of the SLs, through temperature and electric-field reconfiguration of the polar structures. This introduces a previously unexplored degree of freedom in oxide superlattices, with potential implications in thermal management, energy conversion, and other technologies where active control over heat dissipation is critical.

**Acknowledgments**

This work has received financial support from Ministerio de Ciencia (Spain), projects PID2022-138883NB-I00, PID2021-128281NA-I00, TED2021-130930B-I00, Xunta de Galicia (Centro de investigación do Sistema universitario de Galicia accreditation 2023-2027, ED431G 2023/03) and the European Union (European Regional Development Fund -ERDF). The research of F. R. receives financial support from the Oportunius Program, Xunta de Galicia. N. V.-D. acknowledges financial support from MINECO (Spain) through an FPI fellowship (PRE2020-096467). We acknowledge the access granted by the Galician Supercomputing Center (CESGA) to its supercomputing infrastructure. The supercomputer FinisTerrae III and its permanent data storage system have been funded by the NextGeneration EU 2021 Recovery, Transformation and Resilience Plan, ICT2021-006904, and also from the Pluriregional Operational Programme of Spain 2014-2020 of the European Regional Development Fund (ERDF), ICTS-2019-02-CESGA-3, and from the State Programme for the Promotion of Scientific and Technical Research of Excellence of the State Plan for Scientific and Technical Research and Innovation 2013-2016 State subprogramme for scientific and technical infrastructures and equipment of ERDF, CESG15-DE-3114. M.S.C. is the recipient of a Next-Generation EU Maria Zambrano fellowship. A.G.-L. acknowledges financial support through a research grant from the mobility program of Universidad Rey Juan Carlos (Spain). E.L. acknowledges the Serra Húnter Plan from the Generalitat de Catalunya. N.B. acknowledges funding through Advanced Materials programme supported by MCIN with funding from the European Union NextGenerationEU (PRTR-C17.I1) and by the Generalitat de Catalunya. Authors would like to thank the use of the USC Research Infrastructures Area analytical facilities as well as the use of instrumentation provided by the National Facility ELECMI ICTS (ELC568-2025 grant). Authors also acknowledge the use of instrumentation as well as the technical advice provided by the Joint Electron Microscopy Center at ALBA (JEMCA) and funding from Grant IU16-014206 (METCAM-FIB) to ICN2 funded by the European Union through the European Regional Development Fund (ERDF), with the support of the Ministry of Research and Universities, Generalitat de Catalunya. J. S., N.B and X.L. acknowledge B. Mundet, F. Belarre and M. Rosado for the STEM images and FIB fabrication of the lamellas. The ICN2 is funded by the CERCA programme/Generalitat de Catalunya and by the Severo Ochoa Centres of Excellence Programme, funded by the Spanish Research Agency (AEI, CEX2021–001214-S). ICN2 acknowledges funding from Generalitat de Catalunya 2021SGR00457. CBE acknowledges support for this research through a Vannevar Bush Faculty Fellowship (ONR N00014-20-1-2844), and the Gordon and Betty Moore Foundation's EPiQS Initiative, Grant GBMF9065. Thin film synthesis at the University of Wisconsin–Madison was supported by the US Department of Energy (DOE), Office of Science, Office of Basic Energy Sciences (BES), under award number DE-FG02-06ER46327. ICN2 is founding member of e-DREAM.[23]

## METHODS

**Deposition of the samples**. $[(PbTiO_3)_n/(SrTiO_3)_n]_y$ superlattices were epitaxially grown on $(001)_{pc}$ $DyScO_3$ (DSO) and (001) $SrTiO_3$ (STO) substrates by 90º off-axis radiofrequency (RF) magnetron sputtering. The sputtering system was equipped with two independent guns, employing ceramic targets of stoichiometric $SrTiO_3$ and $Pb_{1.2}TiO_3$, the latter containing a 20% excess of Pb to compensate for its volatility during growth. Prior to deposition, the DSO and STO substrates were treated to achieve well-defined -DyO and -$TiO_2$ single-terminated surfaces, respectively, following previously established procedures [24,25].

All the samples were deposited at an RF power of 100 W, under a total pressure of 200 mTorr in a mixed $O_2$/Ar atmosphere with a 9:51 ratio. The deposition temperature was set to 575 ºC for the superlattices grown on STO, while the samples deposited on DSO substrates required a higher temperature of 625 ºC.

The deposition rates of $PbTiO_3$ and $SrTiO_3$ under such conditions were calibrated by ex-situ X-ray reflectivity, yielding ≈0.018 and 0.012 nm/s, respectively, which enabled precise control of superlattice periodicity. The structural quality and periodicity of the superlattices were confirmed by high-resolution X-ray diffraction and HAADF-STEM.

**X-Ray Diffraction analysis**: X-ray diffraction (XRD) measurements were carried out using a high-resolution four-circle laboratory diffractometer (Malvern-Panalytical X'Pert MRD) equipped with a monochromatic Cu $K\alpha_1$ radiation source, selected via a 2×Ge(220) channel-cut monochromator in the primary beam. High-resolution $\omega/2\theta$ scans and reciprocal space maps (RSMs) of selected hkl reflections were acquired using a multi-channel solid-state array detector (PixCel, Malvern-Panalytical). Temperature-dependent measurements were performed from room temperature up to 600 °C using a DHS1100C domed heating chamber (Anton Paar) with a graphite dome. To compensate for potential sample tilts during thermal cycling, RSMs were realigned at each temperature point. Precise determination of the lattice parameters was achieved by analyzing the centroid positions of the measured Bragg reflections. For periodic superstructures, such as superlattice stack or laterally ordered polar domain arrangement, higher-order satellite reflections were used to calculate the average zero-order peak positions. This approach minimized inaccuracies arising from peak overlap with coexisting structural modulations or domain configurations

**TEM analysis**: High-angle annular dark-field scanning transmission electron microscopy (HAADF-STEM) imaging was performed on cross-sectional lamellae of epitaxial $PbTiO_3/SrTiO_3$ (PTO/STO) superlattices. Cross-sectional lamellae were prepared using a Helios 5 UX Focused Ion Beam (FIB) with operating conditions ranging 30 kV down to 2 kV. The HAADF-STEM images were acquired at atomic resolution using a double aberration-corrected STEM microscope (Spectra 300 (S)TEM, ThermoFisher) operated at 200 and 300 kV. High-magnification images were obtained with precise alignment along the [010] zone axis. To ensure sub-pixel accuracy in atomic column position determination, a series of fast STEM images were acquired sequentially and aligned using a drift-corrected frame integration (DCFI) technique based on cross-correlation analysis, and an acquisition method within the Thermo Scientific Velox software. This procedure minimizes scan distortions and enables accurate localization of atomic positions across the full field-of-view of the image. In the resulting HAADF images, atomic columns of Pb, Sr, and Ti are clearly resolved. Following appropriate image filtering (radial Wiener filter), a dedicated python package atomap was employed to accurately extract the centroid positions of the atomic columns.[26] The local displacement of the B-site cations (Ti) was then quantified by comparing their positions with respect to the geometric center of the surrounding four A-site cations (Pb or Sr). These displacements are directly proportional to the local polar distortions in the perovskite lattice. From this analysis, we constructed maps of the polar displacement vectors, represented by arrows whose magnitude and direction correspond to the strength and orientation of the local polarization. These vector maps reveal the spatial distribution of polar displacements across the PTO and STO layers, providing insight into the nanoscale ferroelectric behavior of the heterostructure. Bright field high resolution (HR)TEM and image dark-field TEM images were collected using same microscope and a Ceta™ 16M Camera.

**FDTR Measurements.** The cross-plane thermal conductivity of the films was measured by Frequency Domain Thermoreflectance (FDTR).[27] A sinusoidally modulated pump laser (λ=488 nm, modulating f=2 kHz–50 MHz, spot sizes 1/$e^2$ radius ≈3.7 or 10.5 mm) is focused on the surface of the film, coated by a 100-nm-thick layer of Au, to produce an oscillatory modulation of the surface temperature. This results in a periodic variation of the Au thermoreflectance, which is probed by a laser beam (λ=532 nm). The thermal properties of the sample are obtained by fitting the phase data to an analytical solution of the heat diffusion equation, in a multilayer model.[27] The thermal conductivity of the substrates was obtained from Ref.[28]; for Au we measured the electrical conductivity in co-deposited samples and used the Wiedemann-Franz law to obtain $\kappa_{Au}$. The $C_p$ of the substrate and Au transducer were measured or obtained from the literature.

To separate the effect of the thermal boundary conductance (TBC) on the absolute value of $\kappa$ of the films, they were deposited over half of the substrate while the Au transducer covered the whole surface (film plus bare substrate). Fitting the data from the bare portion of the substrate we obtain the value for the Au/substrate TBC.

# Supplementary information

## *TEM analysis of the superlattices:*

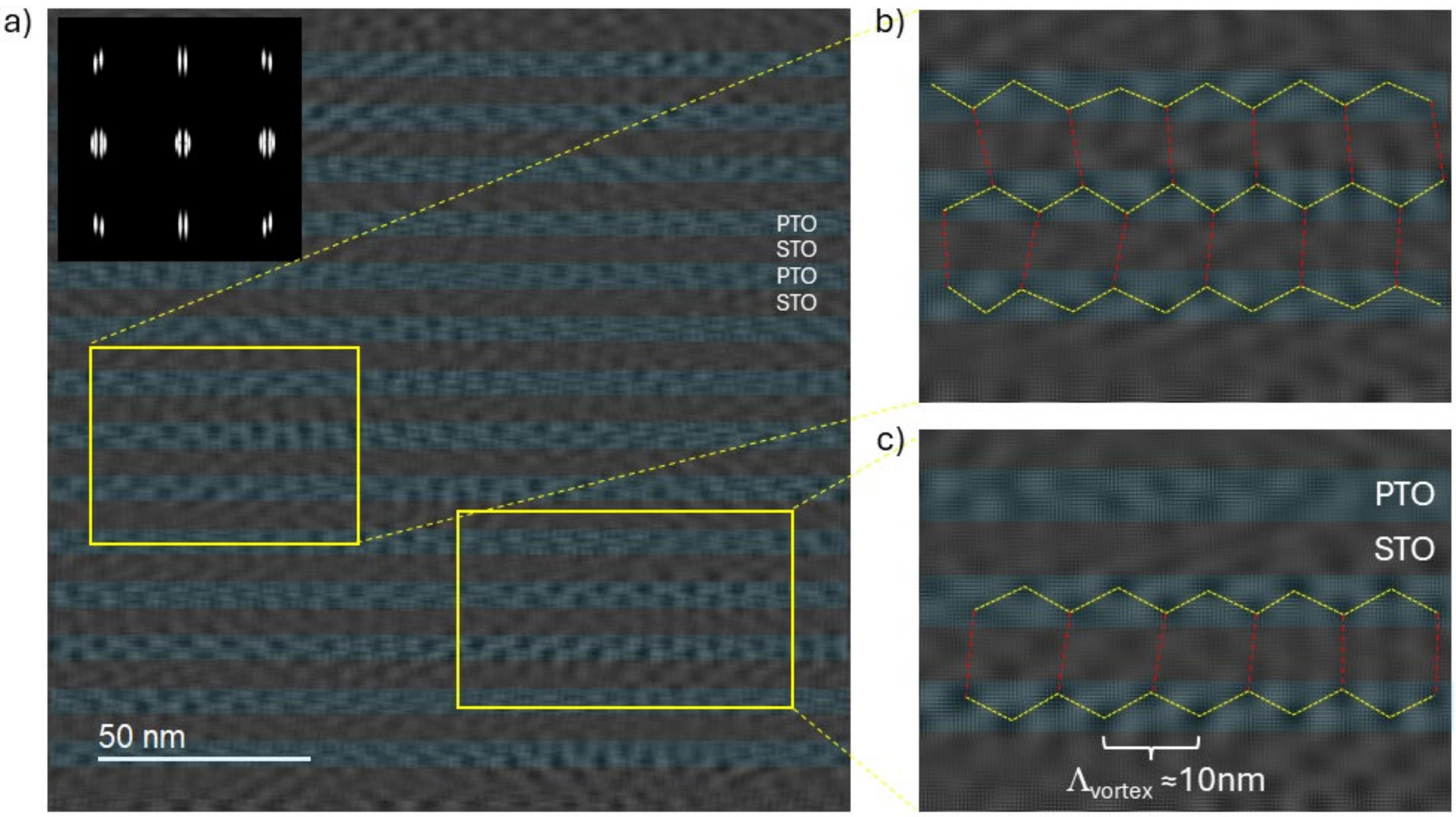

**Figure S1:** a) FFT-filtered HAADF-STEM image of the $[(PTO)_{15}/(STO)_{15}]_{15}$ SL on DSO. The bright layers correspond to PTO, while the dark layers are STO. The contrast within the layers has been enhanced after filtering by applying the mask in the inset (only bright part of the pattern is used for the inverse FFT). The PTO slabs show a checkerboard pattern alternating an up and down wavy arrangement along the horizontal direction. b) and c) enlargement of the previous image depicts two regions where adjacent PTO layers with the wavy pattern seem to couple in antiphase in the vertical direction through the STO layer. In b) the third layer seems again to couple in antiphase, so the vertical periodicity becomes twice the SL PTO/STO bilayer sequence. The lateral periodicity of this arrangement corresponds to 25-26 unit cells, which is about $\Lambda_{vortex} \approx 10$ nm, in agreement with the XRD measurements for the vortex phase.

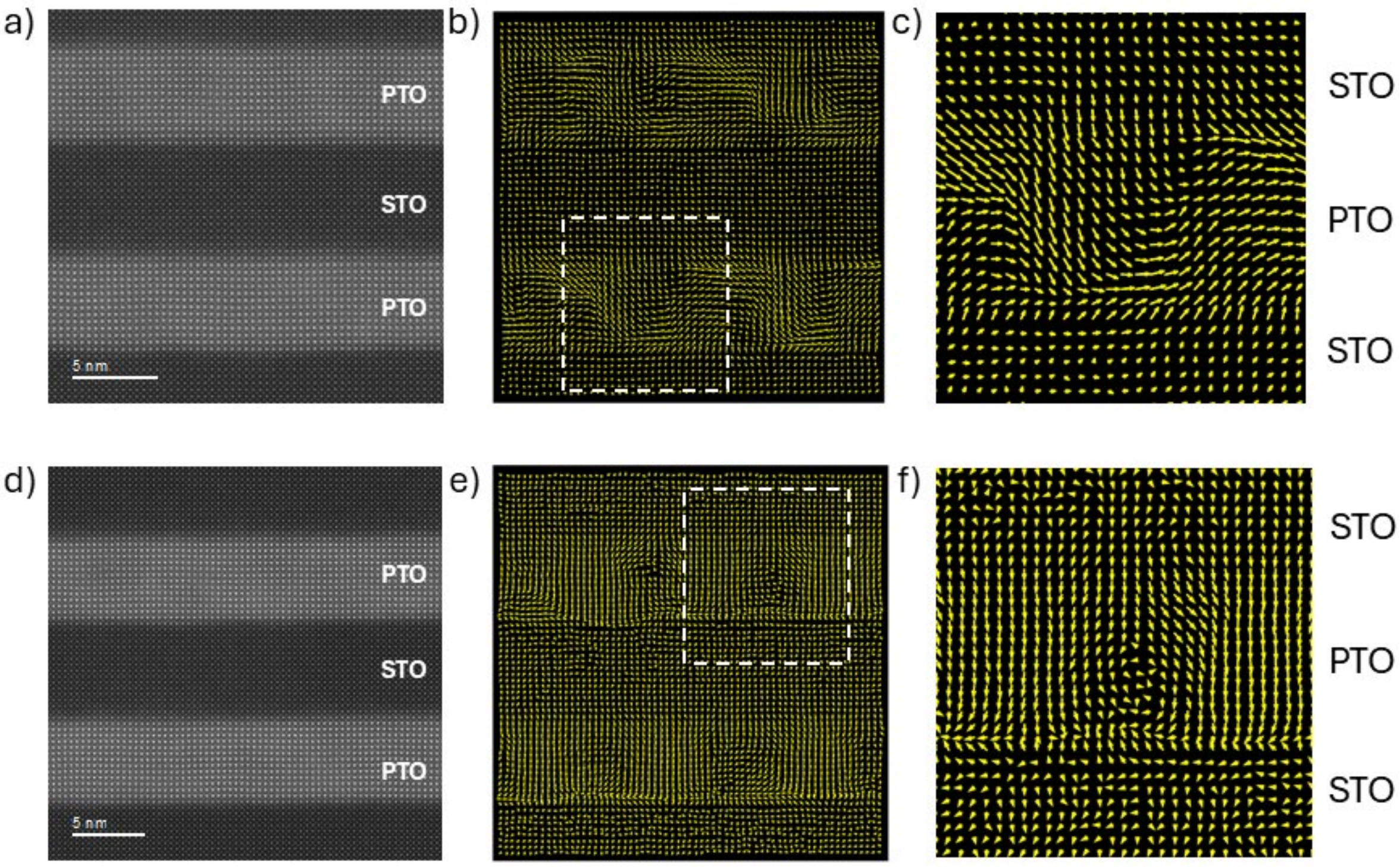

**Figure S2: Comparison between HAADF-STEM observations on the $[(PTO)_{15}/(STO)_{15}]_{15}$ (DSO substrate) cross-section lamella at different conditions.** a) High magnification image (nominal x4M) displaying two consecutive PTO slabs. b) Polar displacement map calculated from previous image (individual dipole vectors indicated with yellow arrows). c) Magnification of the polar map highlighting the wavy polar arrangement along the PTO slab. These images were acquired at 200 kV and are the same as those shown in Figure 1 a-c) in the main manuscript. For the sake of comparison, they are reproduced again along with images taken at similar magnification a few days after at 300 kV (d), and after having thinned the lamellae using FIB with Ga ion gun. The intention was to obtain a higher resolution in the ionic column position. However, the new conditions for the image acquisition resulted in a variation of the polar displacement arrangement in e), more evident in the magnified image (f). Note that in this image the polar displacement consists mostly of downwards *c*-axis oriented polar domains in the PTO layer, approaching single polar domains, periodically disturbed along the horizontal axis by the formation of Néel-type skyrmions, where polarization seems to point towards a localized core. This contrasts with the previously mentioned wave-type arrangement and is proof of the subtle balance between elastic, electrostatic, degrees of freedom controlling the polar topology.

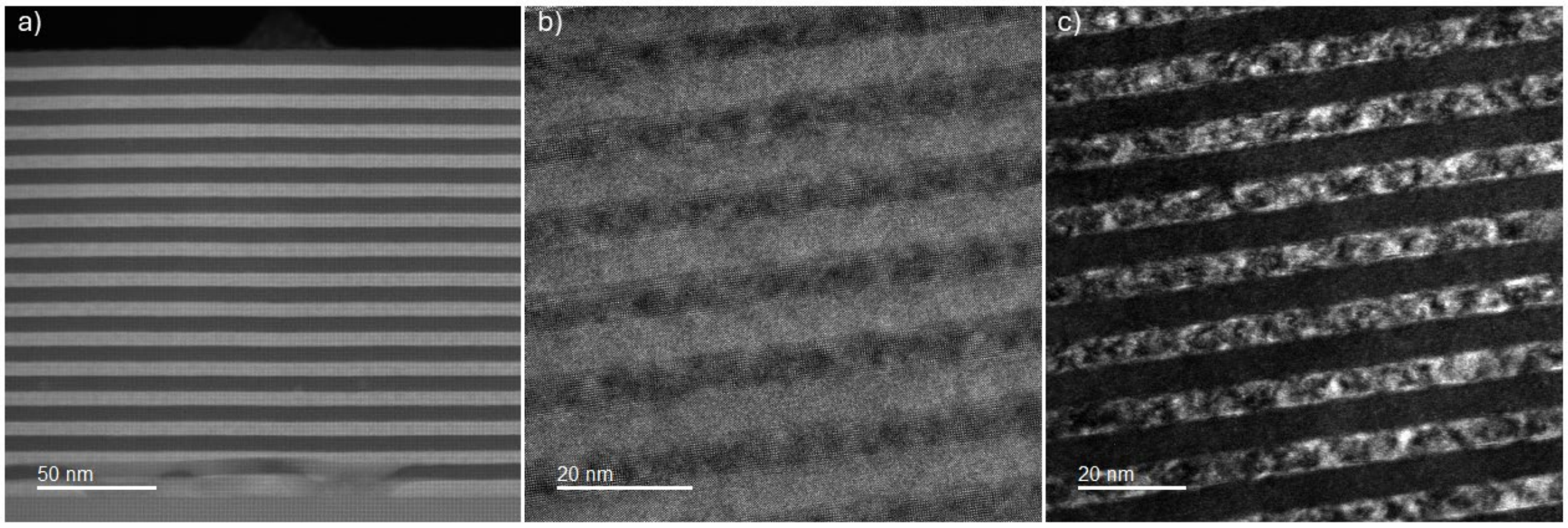

**Figure S3: TEM images obtained with different techniques**. a) HAADF-STEM image of the whole $[(PTO)_{15}/(STO)_{15}]_{15}$ SL cross-section, including the DSO substrate. Note that the first STO and PTO layers in contact with substrate show some growth interruptions, but the second STO/PTO bilayer is already continuous. The SL sequence is perfectly homogenous. The top layer corresponds to STO. b) Bright field HRTEM image shows a contrast between dark layers (PTO) and bright layers (STO). An additional contrast is observed within the PTO layers that corresponds to the lateral periodical arrangement. c) Dark-field TEM image clearly shows the complex periodical contrast across the PTO layers, which corresponds to the polar domain pattern.

## *XRD characterization of the superlattices:*

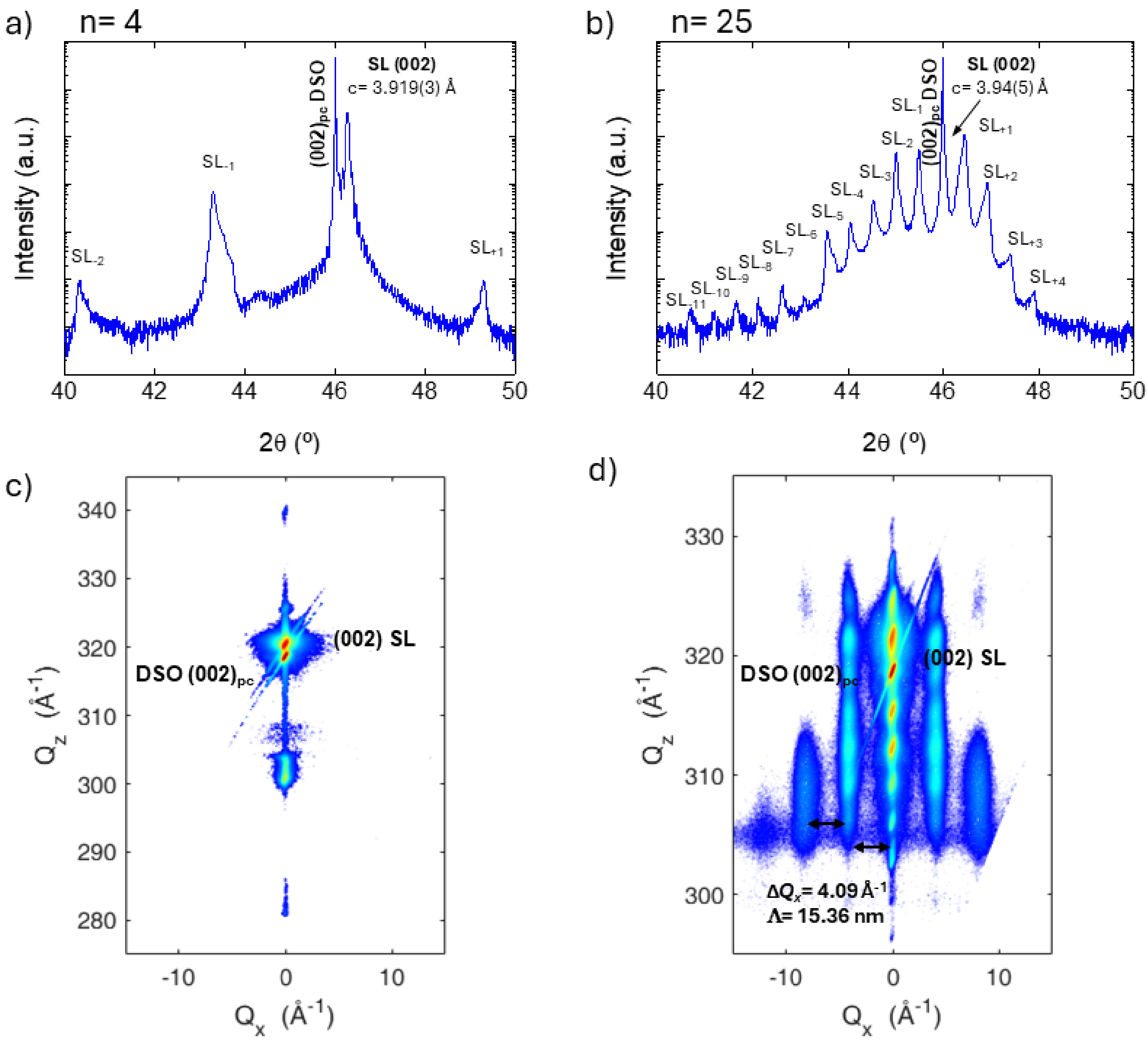

**Figure S4: High-resolution X-ray diffraction characterization of the $[(PTO)_4/(STO)_4]_{60}$ (a,c) and $[(PTO)_{25}/(STO)_{25}]_9$ (b,d) SLs grown on $(001)_{pc}$ DSO**. ω/2θ scans of both n= 4 and n= 25 SLs show well-defined satellite peaks around the $(002)_{pc}$ substrate reflection, indicating high structural coherence and precise bilayer periodicities of $\Lambda_{SL}$= 3.24 and 19.86 nm, respectively. The SL satellite peaks are also visible in the reciprocal space maps shown in (c,d). For n= 25, apart from these satellites, additional lateral signals appear at finite $Q_x$, revealing in-plane structural modulation with a periodicity of $\Lambda_x$≈ 15.36 nm, consistent with the emergence of flux-closure domains

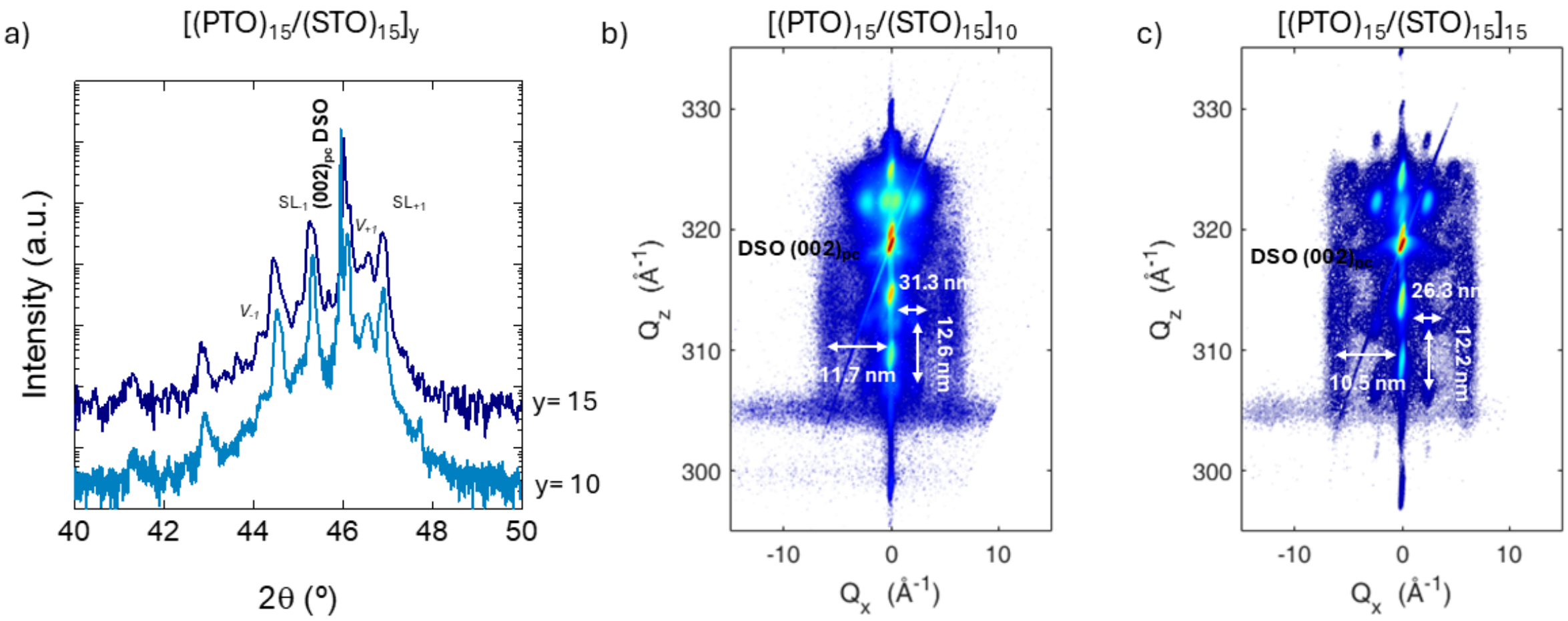

**Figure S5: High-resolution X-ray diffraction characterization of $[(PTO)_{15}/(STO)_{15}]_y$ SLs (y= 10, 15) grown on $(001)_{pc}$ DSO.** $\omega/2\theta$ scans around the $(002)_{pc}$ DSO reflection show well-defined superlattice satellite peaks, along with additional low-intensity reflections, labelled as $V_{\pm n}$, which appear between the satellites and are attributed to periodic modulations due to polar vortex formation. These reflections are better appreciated in the thicker y= 15 sample. The reciprocal space maps exhibit diffuse lateral satellites at finite $Q_x$ positions, revealing the presence of in-plane modulations consistent with the formation of ordered polar vortex arrays. The corresponding lateral periodicities are $\Lambda_{vortex}$=11.70 nm (y= 10) and 10.5 nm (y= 15). Moreover, a second family of satellites emerges at intermediate $Q_x$ and $Q_z$ positions, indicating additional structural modulations with both in-plane and out-of-plane periodicities, as indicated in the image. These features are attributed to the formation of a three-dimensional polar supercrystal (SC) phase, characterized by the long-range periodic arrangement of vortex structures in both lateral and vertical directions.

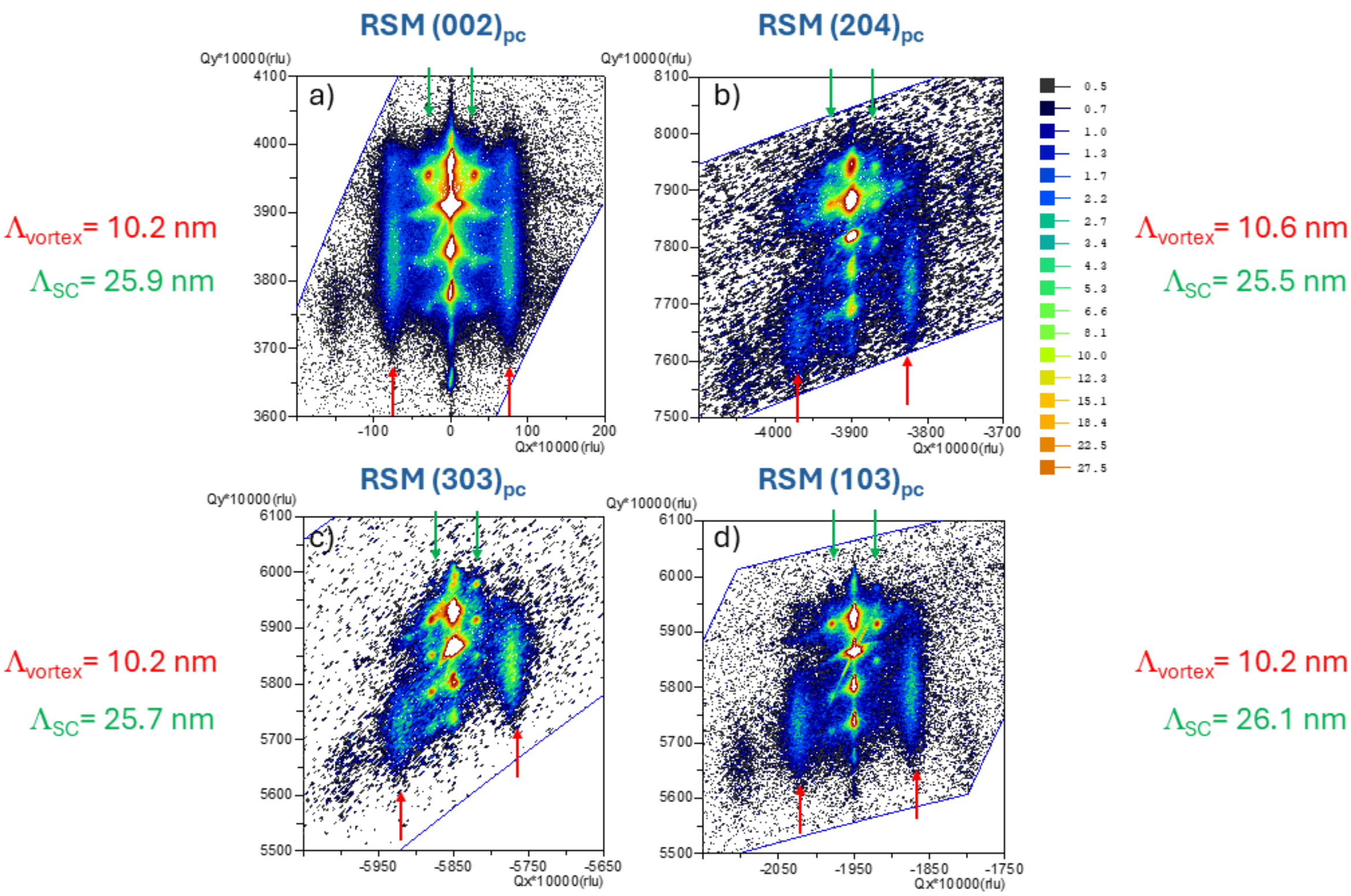

**Figure S6: Reciprocal space maps at room temperature around different substrate reflections for the $[(PTO)_{15}/(STO)_{15}]_{15}$ SL deposited on $(001)_{pc}$ DSO.** a) $(002)_{pc}$ DSO, b) $(204)_{pc}$, c) $(303)_{pc}$, and d) $(103)_{pc}$. The red arrows indicate the first order lateral satellites of the vortex phase, while the green arrows indicate the subset of first order superstructure satellites coming from the supercrystal phase (SC). The corresponding $Q_x$ spacing between first order satellites is equivalent for all reflections for the vortex and SC satellites, which gave around $\Lambda_{vortex} \approx 10.5$ nm and $\Lambda_{SC} \approx 26$ nm, respectively.

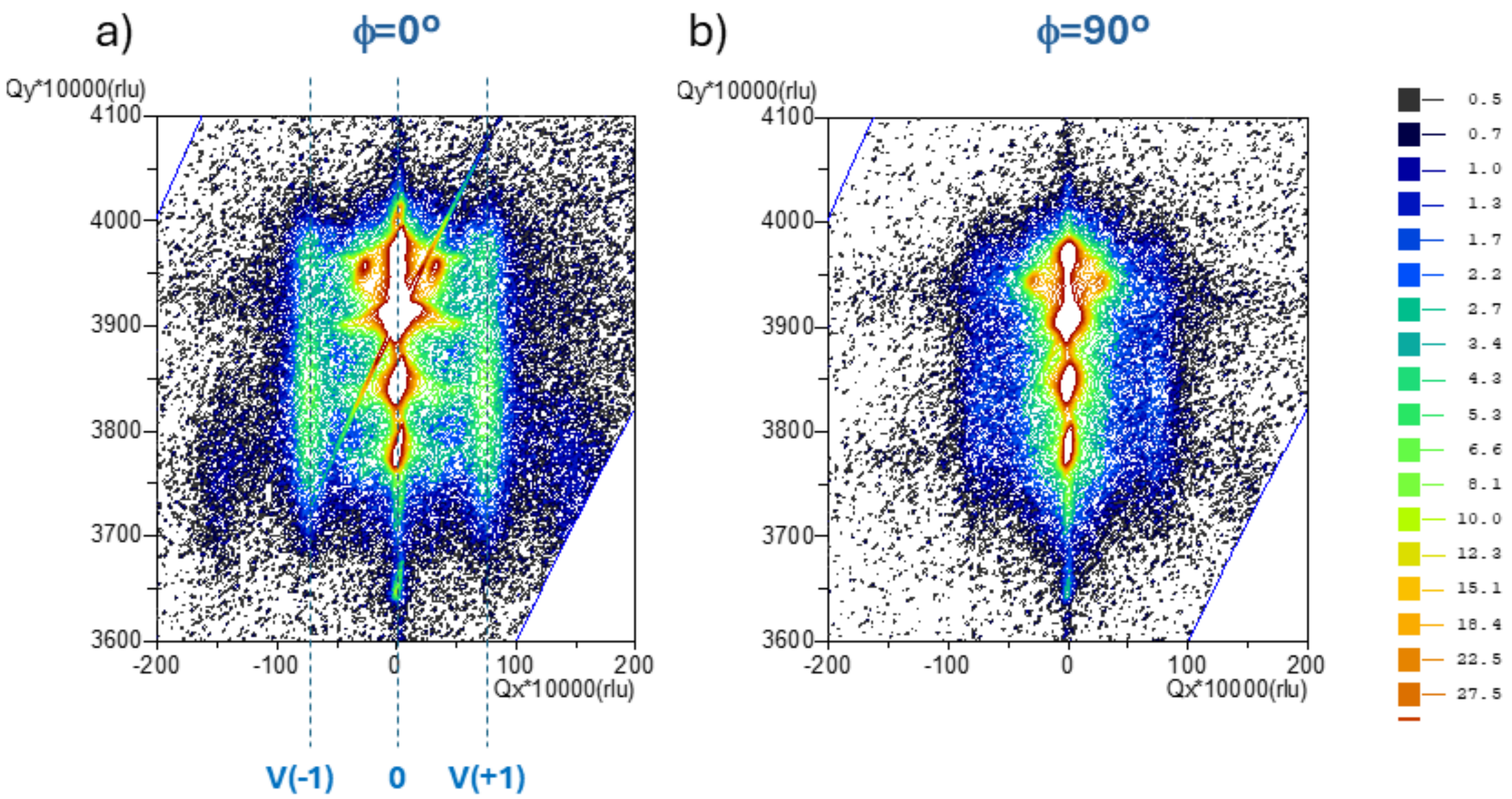

**Figure S7: Reciprocal space maps of $(002)_{pc}$ DSO reflection at room temperature for the $[(PTO)_{15}(STO)_{15}]_{15}$ SL along different in-plane directions.** a) corresponds to azimuthal angle $\phi$=0º // $[001]_O$; and b) $\phi$=90º $\perp$ $[001]_O$ (subindex O: indicates the crystallographic directions of the DSO orthorhombic phase). The vortex periodicity that was clearly observed along the $[001]_O$ directions is not observed in the perpendicular direction. This indicates that vortex domains form stripes aligned perpendicular to $[001]_O$ substrate direction. A different situation is observed for the SC satellites that are equally observed in both crossed directions. This is an indication of the 3D planar arrangement of these domains (at least in the horizontal plane).

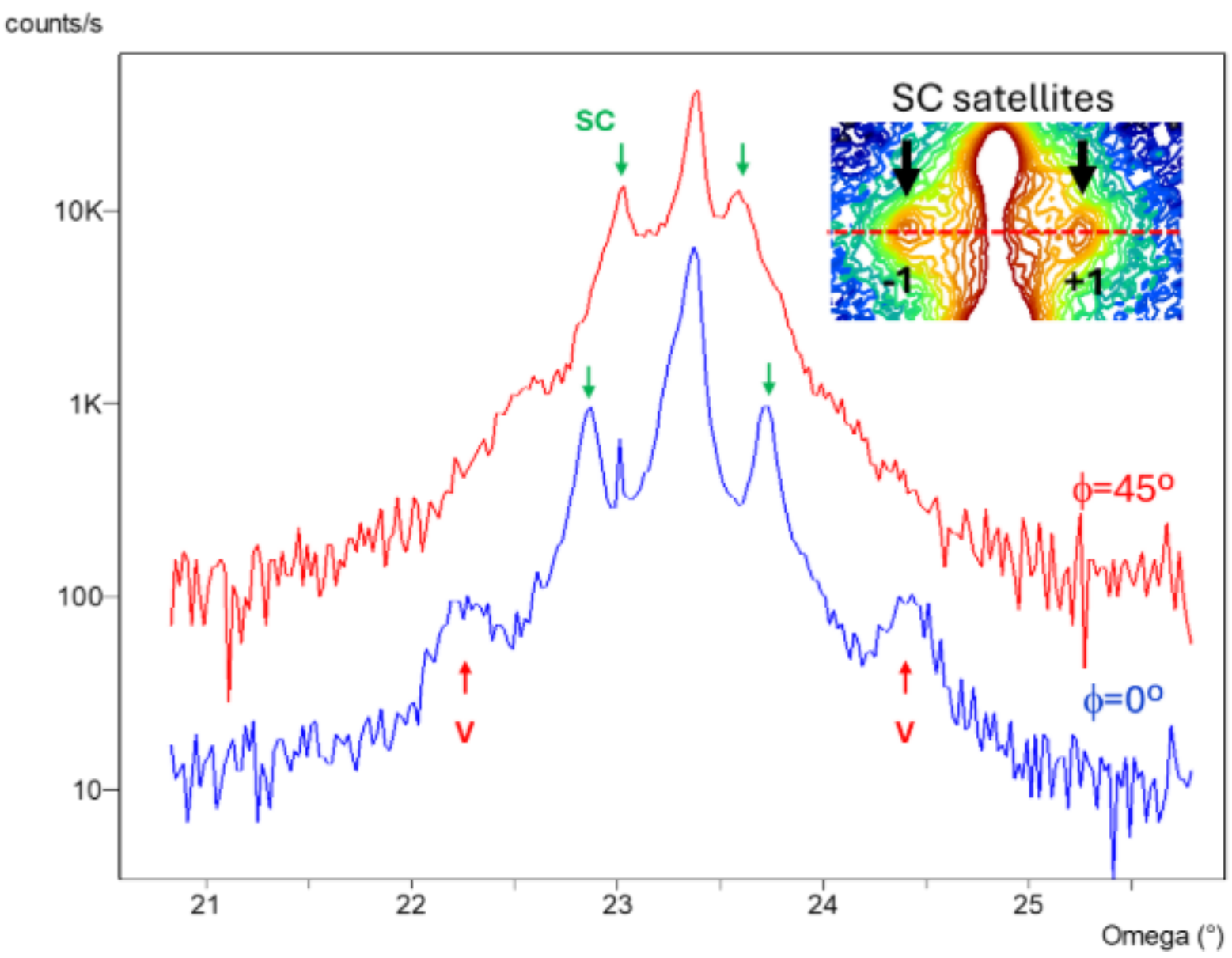

**Figure S8: Omega scans performed on the reciprocal space maps of $(002)_{pc}$ DSO reflection at the $Q_z$ position of the main SC peaks.** These scans were performed along two different in-plane directions either a) at $\phi$=0º // $[100]_{pc}$; and b) $\phi$=45º // $[110]_{pc}$. The SC periodicity can be extracted from the $\Delta\omega$ spacing. Along the $[100]_{pc}$ direction the periodicity is $\Lambda_{100}\approx$ 25.8 nm, while along $[110]_{pc}$ direction is $\Lambda_{110}\approx$ 39.7 nm. This is an indication that the satellites are aligned along $[100]_{pc}$ rather than along $[110]_{pc}$, and rules out the existence of the classical $a_1/a_2$ stripe domains with $[110]_{pc}$ domain walls, typical of *a*-axis oriented PTO films, and points towards this new type of SC arrangement.

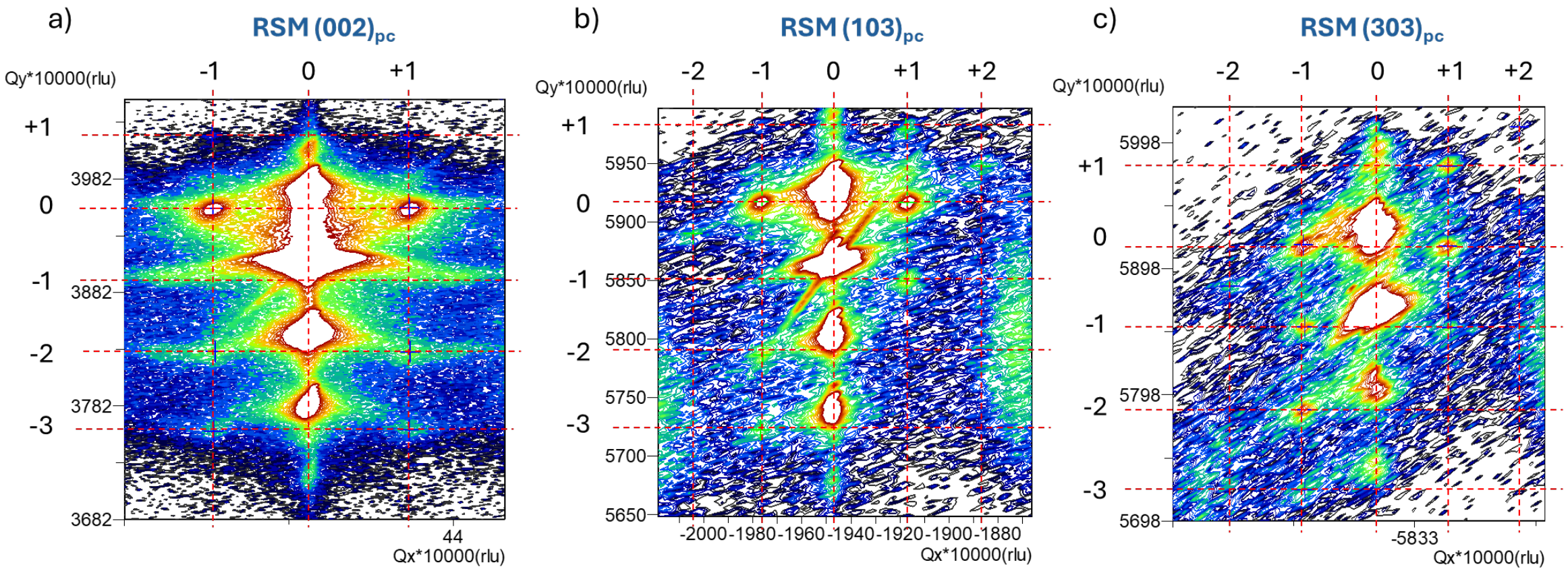

**Figure S9: Subset of satellite positions for the supercrystal (SC) phase in the RSMs for reflections**: a) $(002)_{pc}$; b) $(103)_{pc}$ and c) $(303)_{pc}$ at room temperature for $[(PTO)_{15}/(STO)_{15}]_{15}$ SL on DSO at $\phi$=0º. The vertical and horizontal red lines mark the positions of the satellite array.

**Table S1**. Experimentally determined ($Q_x$,$Q_z$) positions and XRD Intensity for the observed satellites from the SC phase in the RSMs in Figure S9. The zero order peaks (*h*=0) of the SC are not included in this table because the determination is difficult due to the strong overlap with the main ML satellite along the central vertical rods. Note that there are some satellites at half order ±*l* positions, which indicates that in fact the periodicity of the SC phase could be double that of the SL along the vertical direction.

| HKL | Satellite order | | $Q_x$ ($10^{-4}$ r.l.u.) | $Q_z$ ($10^{-4}$ r.l.u.) | Intensity (cps) |
|---|---|---|---|---|---|
| | ±*h* | ±*l* | | | |
| $(002)_{pc}$ | -1 | +1 | -29.38 | 4020.31 | 1.2 |
| | +1 | +1 | 29.81 | 4018.47 | 1.2 |
| | +1 | 0 | 30.55 | 3956.73 | 52.4 |
| | -1 | 0 | -29.06 | 3956.19 | 50.2 |
| | -1 | -2 | -28.50 | 3830.22 | 5.2 |
| | +1 | -2 | 30.75 | 3829.89 | 4.8 |
| | -1 | -3 | -29.00 | 3760.02 | 2.5 |
| | +1 | -3 | 30.53 | 3760.98 | 2.8 |
| $(103)_{pc}$ | +2 | +0.5 | -1917.69 | 5979.63 | 2.4 |
| | +1 | +1 | -1886.74 | 5947.77 | 1.9 |
| | -1 | 0 | -1976.24 | 5917.34 | 12.2 |
| | +1 | 0 | -1917.14 | 5916.86 | 14.7 |
| | -2 | -0.5 | -2004.40 | 5890.09 | 1.9 |
| | +1 | -1 | -1917.30 | 5849.48 | 4.2 |
| | -1 | -1 | -1976.06 | 5786.76 | 3.8 |
| | -1 | -3 | -1976.89 | 5722.42 | 2.4 |
| $(303)_{pc}$ | +2 | +1.5 | -5790.07 | 6010.81 | 0.1 |
| | +1 | +1 | -5819.59 | 5979.72 | 3.3 |
| | +2 | +0.5 | -5789.27 | 5945.62 | 0.4 |
| | -1 | 0 | -5879.47 | 5916.99 | 6.4 |
| | +1 | 0 | -5819.82 | 5915.84 | 3.6 |
| | -2 | -0.5 | -5909.65 | 5884.16 | 1.2 |
| | -1 | -1 | -5878.71 | 5851.81 | 3.2 |
| | +1 | -1 | -5818.43 | 5851.31 | 2.1 |
| | -1 | -2 | -5879.96 | 5786.02 | 4.2 |
| | -2 | -2 | -5878.98 | 5722.67 | 1.1 |

**Table S2**. Determined in-plane (ip), out-of-plane (oop) cell parameters, and vertical and horizontal periodicities of the SC phase from the different analyzed HKL reflections.

| HKL | ip (Å) | oop (Å) | $\Lambda_x$ (nm) | $\Lambda_z$ (nm) |
|---|---|---|---|---|
| $(002)_{pc}$ | - | 3.894 | 25.9 | 11.96 |
| $(103)_{pc}$ | 3.957 | 3.906 | 26.1 | 11.92 |
| $(303)_{pc}$ | 3.951 | 3.907 | 25.7 | 12.00 |
| $(204)_{pc}$ | 3.952 | 3.913 | 25.5 | 12.04 |

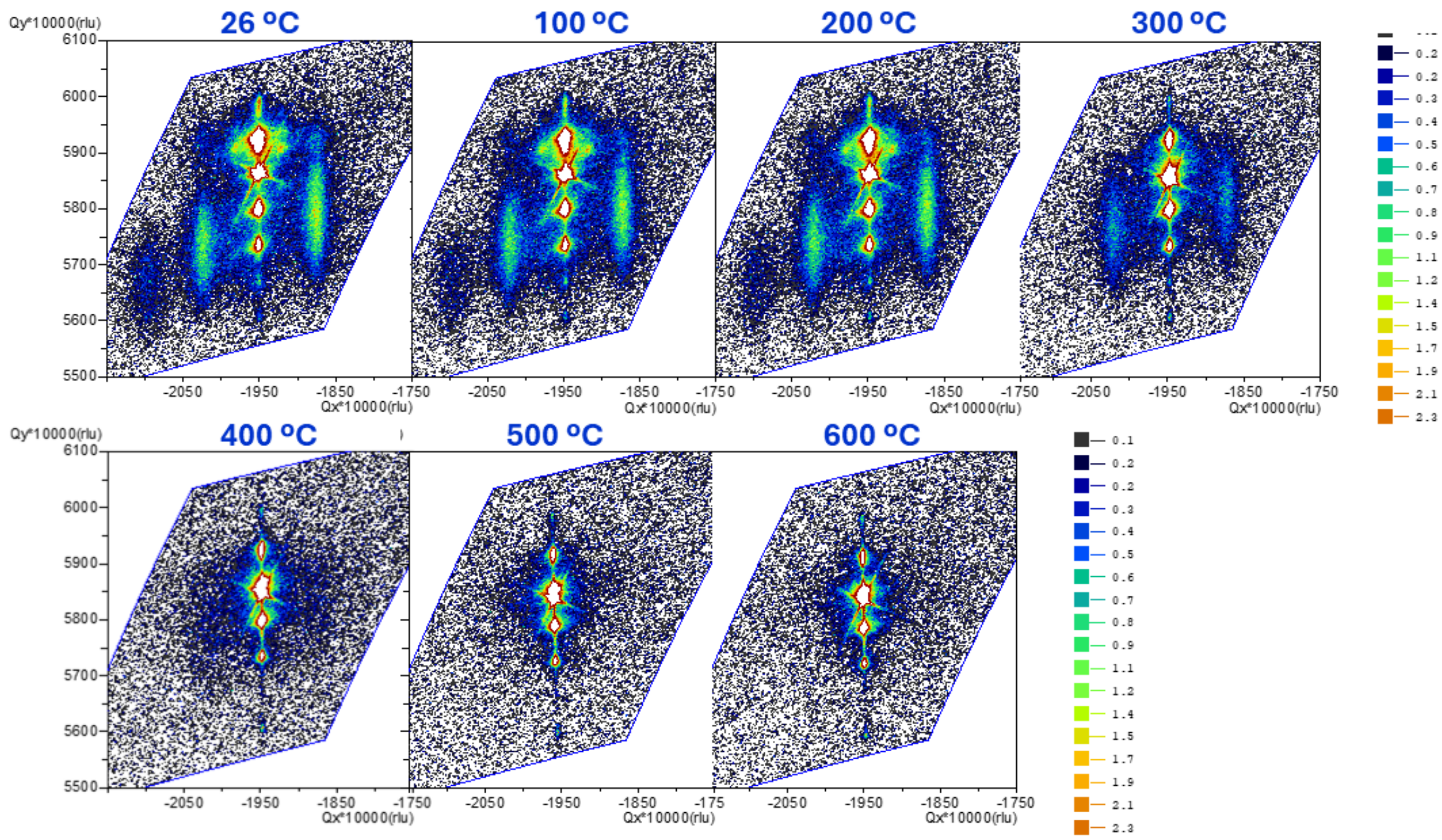

**Figure S10:** Complete set of RSMs measured around the $(103)_{pc}$ reflection of DSO for the $[(PTO)_{15}/(STO)_{15}]_{15}$ SL at different temperatures during a heating cycle from room temperature to 600 ºC.

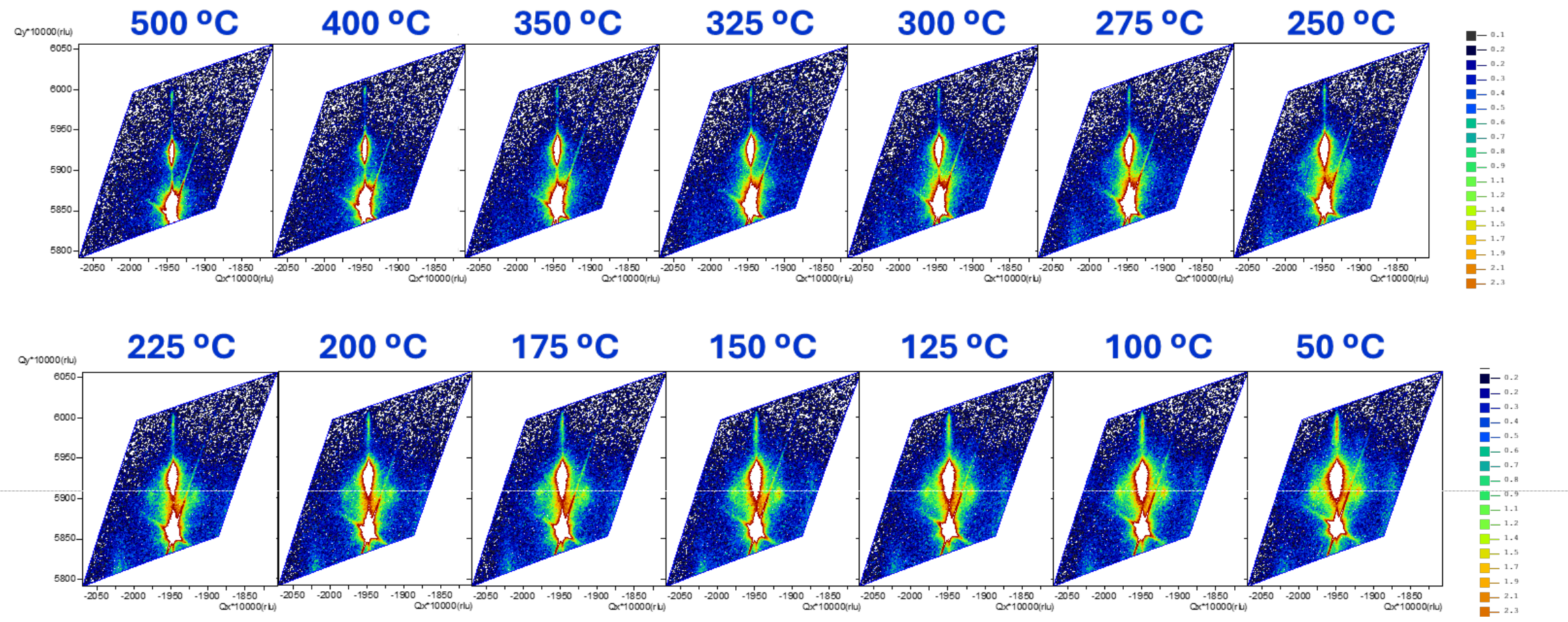

**Figure S11:** Complete set of reduced RSMs measured around the $(103)_{pc}$ of DSO measured for the $[(PTO)_{15}/(STO)_{15}]_{15}$ SL at different temperatures during a cooling cycle from 500 ºC to 50 ºC. The reduced area includes the region of the two most intense satellites of the SC phase.

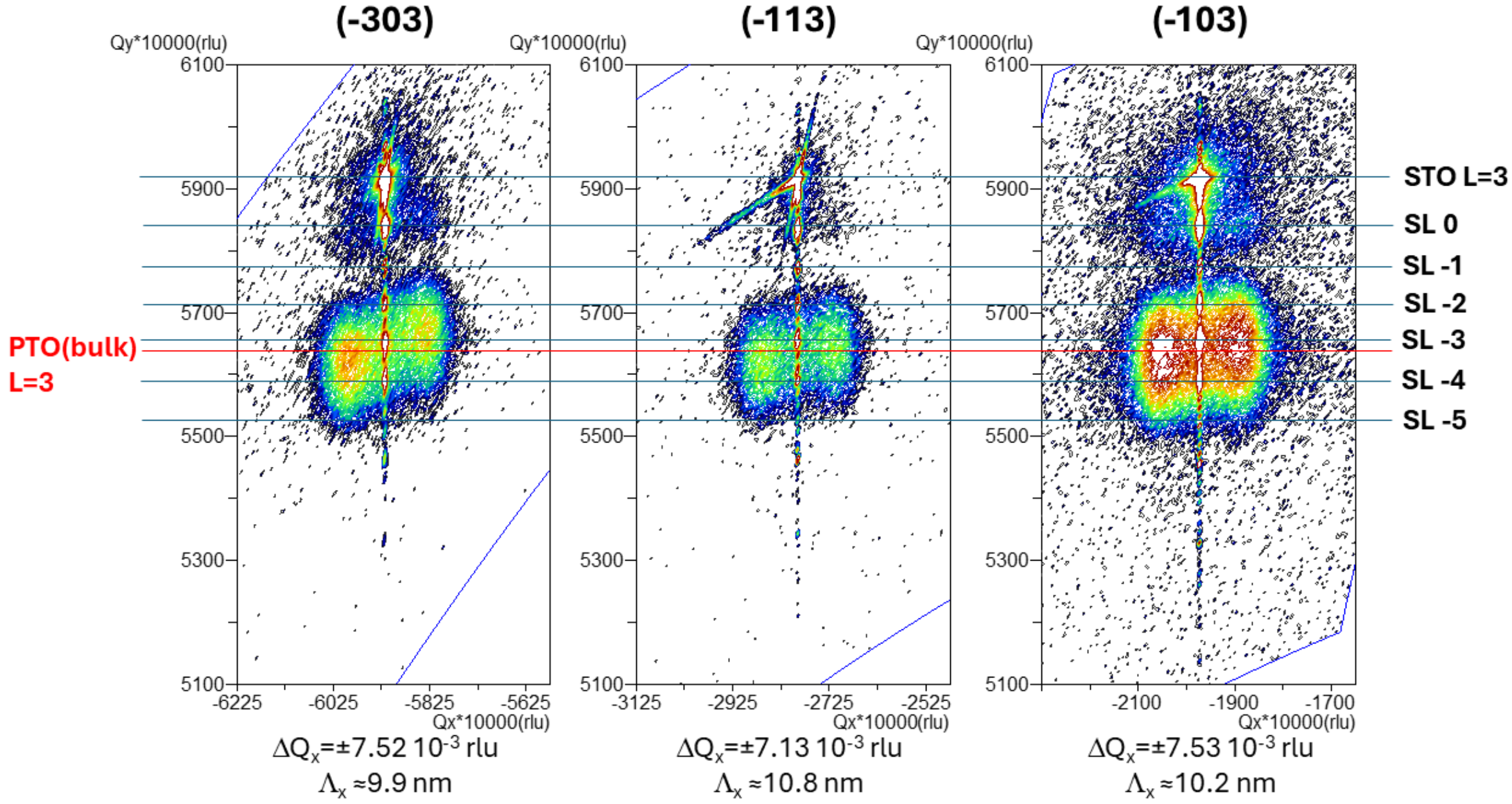

**Figure S12: Reciprocal space maps around the 103, 303 and 113 reflections for the $[(PTO)_{15}/(STO)_{15}]_{15}$ SL deposited on STO (001) substrate.** The horizontal lines indicate the different SL satellite orders, along with the equivalent $Q_z$ position of substrate L=3 peaks. The red horizontal line indicates the average $Q_z$ position of the centroid satellite coming for the lateral vortex arrangement. Note that the left and right centroid $Q_z$ positions are split depending on the HK values, which indicates either a non-vertical component of the domain walls between predominant *c*-axis oriented PTO domains in the PTO layers, or a shear strain in the PTO structure, or maybe the combination of both.

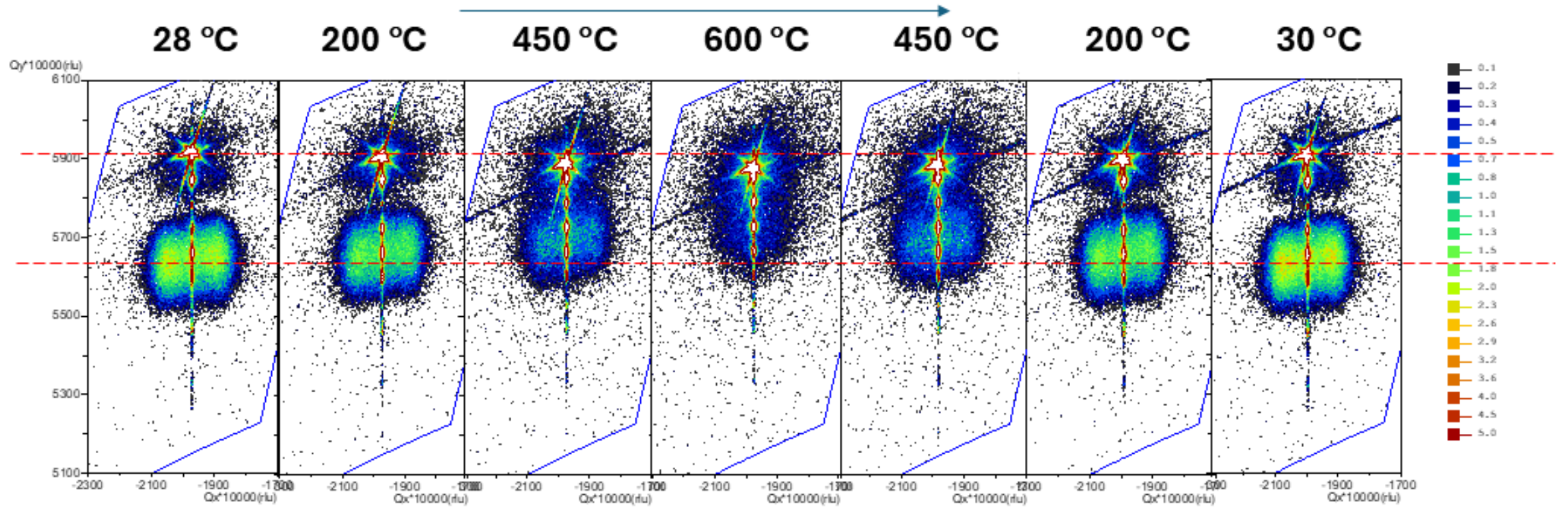

**Figure S13: Reciprocal space maps around the 103 reflection for the $[(PTO)_{15}/(STO)_{15}]_{15}$ SL on STO (001) at different temperatures for heating and cooling cycles.** The red dashed horizontal line indicates the $Q_z$ position of the centroid satellite of the vortex arrangement, as well as that of STO substrate. Note that the $Q_z$ position for the vortex phases increases with temperature (corresponding oop decreases with temperature), contrary to the thermal expansion of STO peak position.

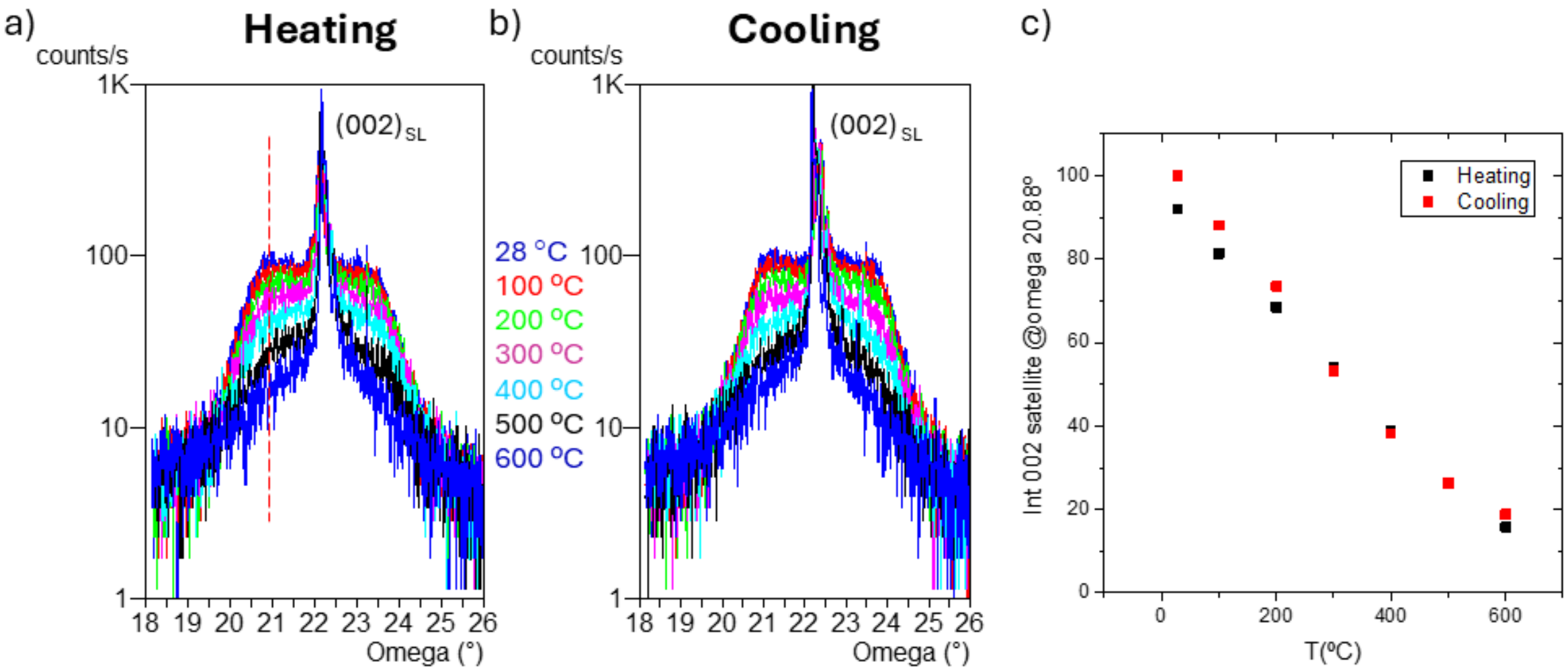

**Figure S14**: **Omega scans performed around the satellite peaks of the 002 reflections of the vortex phase for the $[(PTO)_{15}/(STO)_{15}]_{15}$ SL on STO (001) at different temperatures** (from room temperature to 600ºC) for a) heating and b) cooling cycles. The intensities of the satellites were integrated to obtain the corresponding temperature dependence on panel c).

## ***Temperature dependence of the thermal conductivity:***

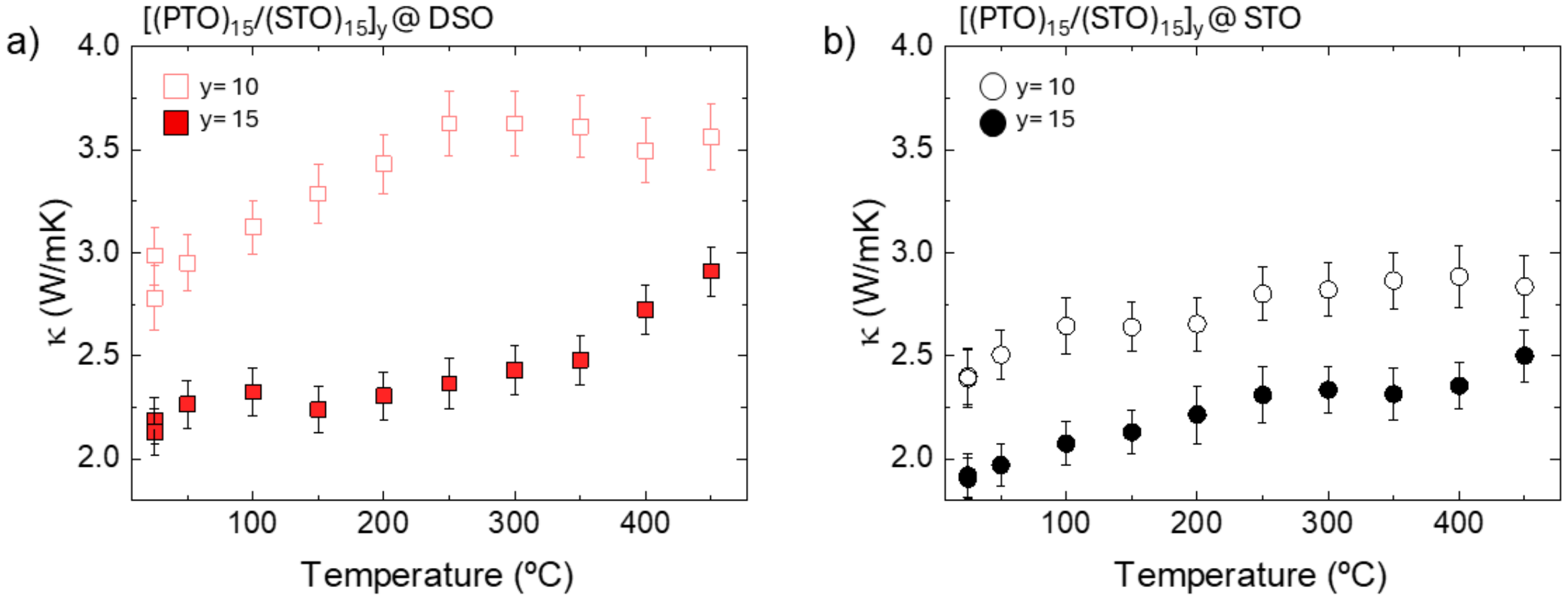

**Figure S15: Temperature-dependent thermal conductivity measurements of $[(PTO)_{15}/(STO)_{15}]_y$ SLs grown on $(001)_{pc}$ DSO (a) and (001) STO (b) substrates.** In both cases, there is a reduction in thermal conductivity when increasing the number of repetitions (y), pointing to localization-like phenomena.

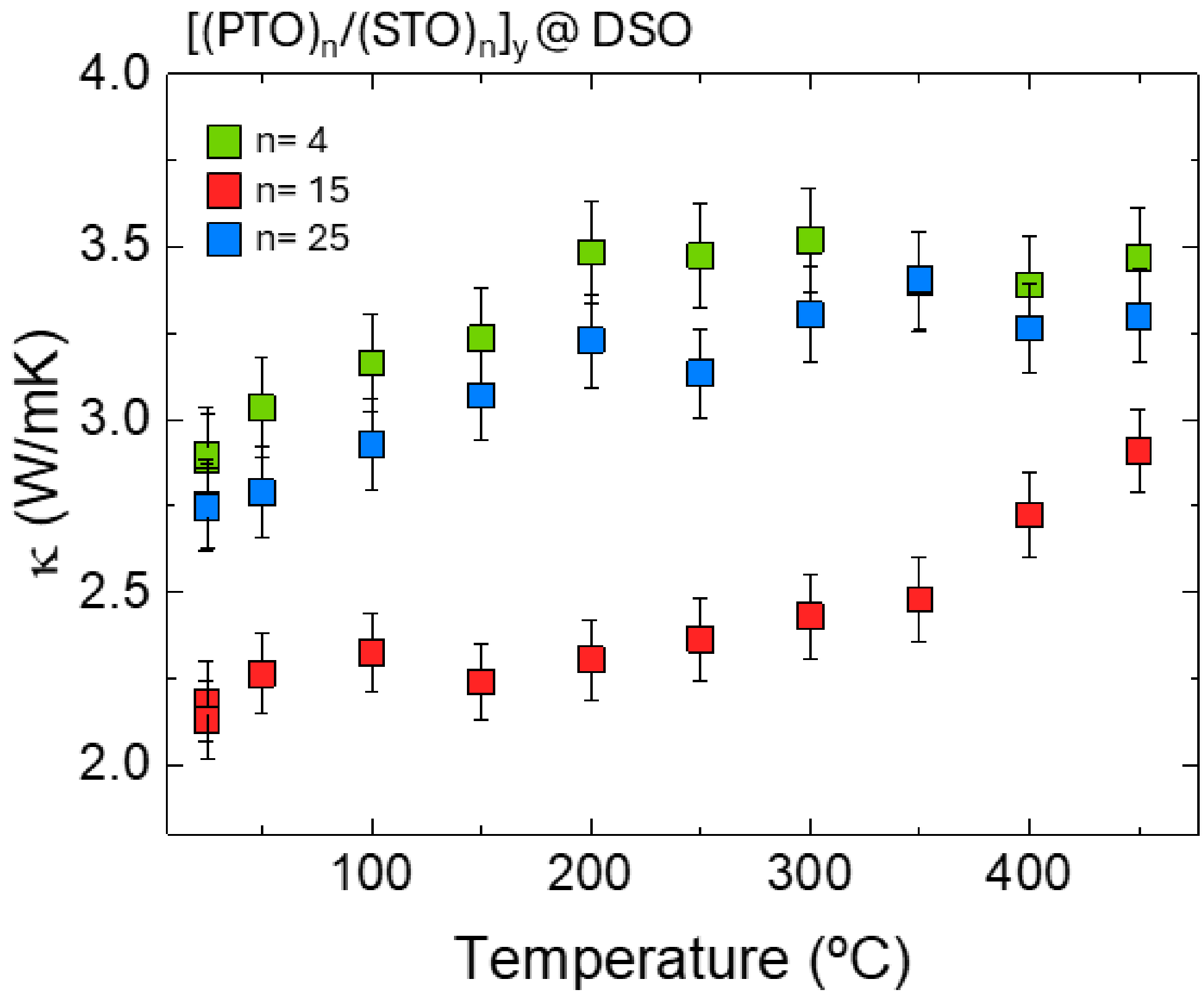

**Figure S16: Temperature-dependent thermal conductivity measurements of $[(PTO)_n/(STO)_n]_y$ superlattices grown on DSO, with the same total thickness (≈170 nm) but different superlattice periodicity (n= 4, 15 and 25 u.c.).** Superlattices with n= 4 and 25 u.c. show the expected trend for nanostructured systems, but the SL with n= 15 shows considerably lower κ, with abrupt changes that are consistent with the transition temperatures of the polar supercrystal and vortices, respectively, as observed in XRD experiments.